\documentclass[journal=jctcce,manuscript=article,layout=traditional]{achemso}
\usepackage{graphicx,dcolumn,bm,xcolor,microtype,hyperref,multirow,amsmath,amssymb,amsfonts,physics,float,lscape,soul,rotating,longtable}
\usepackage[version=4]{mhchem}

\usepackage[normalem]{ulem}

\definecolor{darkgreen}{RGB}{0, 180, 0}

\newcommand{\EFCI}{E_\text{FCI}}

\newcommand{\EsCI}{E_\text{sCI}}
\newcommand{\EPT}{E_\text{PT2}}

\newcommand{\TDDFT}{TD-DFT}
\newcommand{\CASSCF}{CASSCF}
\newcommand{\CASPT}{CASPT2}
\newcommand{\ADC}[1]{ADC(#1)}
\newcommand{\CC}[1]{CC#1}
\newcommand{\CCSD}{CCSD}
\newcommand{\EOMCCSD}{EOM-CCSD}
\newcommand{\CCSDT}{CCSDT}
\newcommand{\CCSDTQ}{CCSDTQ}
\newcommand{\CCSDTQP}{CCSDTQP}
\newcommand{\CI}{CI}
\newcommand{\sCI}{sCI}
\newcommand{\exCI}{exFCI}
\newcommand{\FCI}{FCI}

\newcommand{\AVDZ}{\emph{aug}-cc-pVDZ}
\newcommand{\AVTZ}{\emph{aug}-cc-pVTZ}

\newcommand{\AVQZ}{\emph{aug}-cc-pVQZ}
\newcommand{\DAVQZ}{d-\emph{aug}-cc-pVQZ}
\newcommand{\TAVQZ}{t-\emph{aug}-cc-pVQZ}
\newcommand{\AVPZ}{\emph{aug}-cc-pV5Z}
\newcommand{\DAVPZ}{d-\emph{aug}-cc-pV5Z}
\newcommand{\PopleDZ}{6-31+G(d)}

\newcommand{\IneV}[1]{#1~eV}
\newcommand{\InAU}[1]{#1~a.u.}
\newcommand{\InAA}[1]{#1~\AA}
\newcommand{\MaxP}{Max($+$)}
\newcommand{\MaxN}{Max($-$)}

\newcommand{\pis}{\pi^\star}
\newcommand{\Ryd}{\mathrm{R}}

\newcommand{\SI}{Supporting Information}

\newcommand{\LCPQ}{Laboratoire de Chimie et Physique Quantiques, Universit\'e de Toulouse, CNRS, UPS, France}
\newcommand{\CEISAM}{Laboratoire CEISAM - UMR CNRS 6230, Universit\'e de Nantes, 2 Rue de la Houssini\`ere, BP 92208, 44322 Nantes Cedex 3, France}

\title{A Mountaineering Strategy to Excited States: Highly-Accurate Reference Energies and Benchmarks}

\author{Pierre-Fran{\c c}ois Loos}
	\email{loos@irsamc.ups-tlse.fr}
	\affiliation[LCPQ, Toulouse]{\LCPQ}
\author{Anthony Scemama}
	\affiliation[LCPQ, Toulouse]{\LCPQ}
\author{Aymeric Blondel}
	\affiliation[UN, Nantes]{\CEISAM}    
\author{Yann Garniron}
	\affiliation[LCPQ, Toulouse]{\LCPQ}
\author{Michel Caffarel}
	\affiliation[LCPQ, Toulouse]{\LCPQ}
\author{Denis Jacquemin}
	\email{Denis.Jacquemin@univ-nantes.fr}
	\affiliation[UN, Nantes]{\CEISAM}

\begin{document}
\begin{abstract}
Striving to define very accurate vertical transition energies, we perform both high-level coupled cluster (CC) calculations (up to {\CCSDTQP}) and selected configuration interaction ({\sCI}) calculations (up to several millions of 
determinants) for 18 small compounds (water, hydrogen sulfide, ammonia, hydrogen chloride, dinitrogen, carbon monoxide, acetylene, ethylene, formaldehyde, methanimine, thioformaldehyde, acetaldehyde, cyclopropene, diazomethane, 
formamide, ketene, nitrosomethane and the smallest streptocyanine). By systematically increasing the order of the CC expansion, the number of determinants in the CI expansion as well as the size of the one-electron basis set, 
we have been able to reach near full CI (FCI) quality transition energies. These calculations are carried out on {\CC{3}}/{\AVTZ} geometries, using a series of increasingly 
large atomic basis sets systematically including diffuse functions.  In this way, we define a list of 110 transition energies for states of various characters (valence, Rydberg, $n \rightarrow \pis$, $\pi \rightarrow \pis$,  singlet, triplet, etc.) 
to be used as references for further calculations.  Benchmark transition energies are provided at the {\AVTZ} level as well as with additional basis set corrections, in order to obtain results close to the complete basis set limit. 
These reference data are used to benchmark a series of twelve excited-state wave function methods accounting for double and triple contributions, namely {\ADC{2}}, {\ADC{3}}, CIS(D), CIS(D$_\infty$), {\CC{2}}, STEOM-CCSD,  {\CCSD},  
CCSDR(3), CCSDT-3,  {\CC{3}}, {\CCSDT} and {\CCSDTQ}. It turns out that {\CCSDTQ} yields a negligible difference with the extrapolated {\CI} values with a mean absolute error as small as \IneV{0.01}, whereas the
coupled cluster approaches including iterative triples are also very accurate (mean absolute error of \IneV{0.03}). Consequently, CCSDT-3 and  {\CC{3}} can be used to define reliable benchmarks. This observation does not hold 
for {\ADC{3}} that delivers quite large errors for this set of small compounds, with a clear tendency to overcorrect its second-order version, {\ADC{2}}. Finally, we discuss the possibility to use basis set extrapolation approaches
so as to tackle more easily larger compounds.
\end{abstract}
\clearpage

%
%
\section{Introduction}

Defining an effective method reliably providing accurate excited-state energies and properties remains a major challenge in theoretical chemistry.  For practical applications, the most popular approaches are the complete active 
space self-consistent field ({\CASSCF}) \cite{Heg79,Tay84} and  the time-dependent density functional theory ({\TDDFT})\cite{Cas12,Ulr12b} methods for systems dominated by static and dynamic electron correlation 
effects, respectively. When these schemes are not sufficiently accurate, one often uses methods including second-order perturbative corrections. For {\CASSCF}, a natural choice is {\CASPT}, \cite{And90} but this method 
rapidly becomes impractical for large compounds. If a single-reference method is sufficient, the most popular second-order approaches are probably the second-order algebraic diagrammatic construction, {\ADC{2}}, \cite{Dre15} 
and the second-order coupled cluster, {\CC{2}}, 
methods, \cite{Chr95,Hat00} that both offer an attractive $\order*{N^5}$ scaling (where $N$ is the number of basis functions) allowing applications up to systems comprising ca.~100 atoms.  Compared to {\TDDFT},\cite{Lau14} these 
approaches have the indisputable advantage of being free of the choice of a specific exchange-correlation functional. Using {\ADC{2}} or {\CC{2}} generally provides more systematic errors with respect to reference values than 
TD-DFT, although the improvements in terms of error magnitude are often rather moderate (at least for valence singlet states). \cite{Win13,Jac15b,Oru16} Importantly, both {\ADC{$n$}} and {\CC{$n$}} offer a systematic pathway 
for improvement via an increase of the expansion order $n$. For example, using {\CCSD}, {\CCSDT}, {\CCSDTQ}, etc., allows to check the quality of the obtained estimates. However, in practice, one can only contemplate such 
systematic approach and the ultimate choice of a method for excited-state calculations is often guided by previous benchmarks. These benchmark studies are either performed using experimental or theoretical reference values. 
While the former approach allows in principle to rely on an almost infinite pool of reference data, most measurements are performed in solution and provide absorption bands that can be compared to theory only with the use of 
extra approximations for modeling environmental and vibronic effects.  {In addition, the most accurate experimental data are obtained for 0-0 energies, whereas obtaining trustworthy experimental estimates of vertical transition 
energies is an extremely difficult task, generally requiring to back-transform spectroscopic vibronic data through a numerical process,} \cite{Odd85} {an approach that is typically only applicable to diatomics.}
Consequently, it is easier to use first-principle reference values as benchmarks, as they allow to assess theoretical methods more consistently (vertical values, 
same geometries, no environmental effects, etc). This is well illustrated by the recent contribution of Schwabe and Goerigk, \cite{Sch17} who decided to compute third-order response CC ({\CC{3}})\cite{Chr95b,Koc97} reference 
values instead of using the previously collected experimental values for the test set originally proposed by Gordon's group. \cite{Lea12}

Whilst many benchmark sets have been proposed for excited states, \cite{Par02,Die04b,Gri04b,Rhe07,Pea08,Jac08b,Jac09c,Goe09,Car10,Lea12,Jac12d,Ise12,Win13,Jac15b,Hoy16} the most praised database of theoretical excited 
state energies is undoubtedly the one set up by Thiel and his co-workers. In 2008, they proposed a large set of theoretical best estimates (TBE) for 28 small and medium CNOH organic compounds. \cite{Sch08}  
More precisely, using some literature values but mainly their own {\CC{3}}/TZVP and {\CASPT}/TZVP results computed on MP2/6-31G(d) geometries, these authors determined 104 singlet and 63 triplet reference 
excitation energies.  The same group soon proposed {\AVTZ} TBE for the same set of compounds, \cite{Sil10b,Sil10c} though some {\CC{3}}/{\AVTZ} reference values were estimated by a basis set extrapolation technique. 
In their conclusion, they stated that they ``\emph{expect this benchmark set to be useful for validation and development purposes, and anticipate future improvements and extensions of this set through further 
high-level calculations}''.\cite{Sch08}  The first prediction was soon realized. Indeed, both the TZVP and {\AVTZ} TBE were applied to benchmark various computationally-effective methods, including semi-empirical approaches, 
\cite{Sil10,Dom13,Voi14} {\TDDFT}, \cite{Sil08,Goe09,Jac09c,Roh09,Jac10c,Jac10g,Mar11,Jac11a,Hui11,Del11,Tra11,Pev12,Mai16} the second-order polarization propagator approximation (SOPPA), \cite{Sau15} {\ADC{2}}, 
\cite{Har14} the second order $N$-electron valence perturbation theory (NEVPT2), \cite{Sch13b} the random phase approximation (RPA), \cite{Yan14b} as well as several {\CC{}} variants. \cite{Sau09,Dem13,Pie15,Taj16,Ris17,Dut18} 
In contrast, even a decade after the original work appeared, the progresses aiming at improving and/or extending Thiel's set have been much less numerous. To the best of our knowledge, these extensions are limited to the more 
compact TZVP basis set, \cite{Wat13,Dem13,Har14,Kan14} but in one case. \cite{Kan17} This diffuse-less basis set offers clear computational advantages and avoids some state mixing. However, it has a clear tendency to 
overestimate transition energies, especially for Rydberg states, and it makes comparisons between methods more difficult as basis set dependencies are significantly different in wave function-based and density-based methods. \cite{Lau15} 

Let us now briefly review these efforts. In 2013, Watson \emph{et al.}~obtained with the TZVP basis set and {\CCSDT}-3 --- a method employing an iterative approximation of the triples --- transition energies very similar to the {\CC{3}} 
values. \cite{Wat13}  Nevertheless, as noted the same year by Nooijen and coworkers who also reported  {\CCSDT}-3/TZVP values, \cite{Dem13} ``\emph{the relative accuracy of EOM-CCSDT-3 versus CC3 compared to full CI 
(or EOM-CCSDT) is not well established}''. In 2014,  Dreuw and co-workers performed {\ADC{3}} calculations on Thiel's set and concluded that ``\emph{based on the quality of the existing benchmark set it 
is practically not possible to judge whether {\ADC{3}} or {\CC{3}} is more accurate}''. The same year, Kannar and Szalay, revisited Thiel's set and proposed {\CCSDT}/TZVP reference energies for 17 singlet states of six molecules. 
\cite{Kan14} Recently the same group reported  {\CCSDT}/{\AVTZ} transition energies for valence and Rydberg states of five compact molecules, \cite{Kan17} and used these values to benchmark several 
simpler {\CC{}} approaches.  To the best of our knowledge, these stand as the highest-level values reported to date.  However, it remains difficult to know if these {\CCSDT} transition energies are significantly more accurate 
than their {\CC{3}},  {\CCSDT}-3 or {\ADC{3}}  counterparts. Indeed, for the $\pi \rightarrow \pis$ valence singlet excited state of ethylene, the {\CC{3}}/TZVP, {\CCSDT}/TZVP and {\CCSDTQ}/TZVP estimates of 
\IneV{$8.37$}, \IneV{$8.38$}, and \IneV{$8.36$} (respectively) are nearly identical. \cite{Kan14}

Herein, we propose to continue the quest for ultra-accurate excited-state reference energies. First, although this prevents direct comparisons with previously-published data, we decided to use more accurate {\CC{3}}/{\AVTZ} 
geometries for all the compounds considered here. Second, we employ only diffuse-containing Dunning basis sets to be reasonably close from the complete basis set limit. Third, we climb the mountain via two faces following: 
i) the {\CC{}} route (up to the highest computationally possible order), and ii) the configuration interaction ({\CI}) route with the help of selected {\CI} ({\sCI}) methods. By comparing the results of these two approaches, it is 
possible to get some reliable information about how far our results are from the full CI  ({\FCI}) ones.  Fourth, in order not to limit our investigation to vertical absorption, we also report, in a few cases, fluorescence energies. 
 Of course, such extreme choices impose drastic restrictions on the size of the molecules one can treat with such approaches. However, we claim here that they allow to accurately estimate the {\FCI} result for most excited states.

%
%
\section{Computational Details}
\label{sec-met}

\subsection{Geometries}

All geometries are obtained at the {\CC{3}}/{\AVTZ} level without applying the frozen core approximation. These geometries are available in the {\SI} (SI). While several structures are extracted from Ref.~\citenum{Bud17} 
(acetylene, diazomethane, ethylene, formaldehyde, ketene, nitrosomethane, thioformaldehyde and streptocyanine-C1) , additional optimizations are performed here following the same protocol as in that earlier work. 
First, we optimize the structures and compute the vibrational spectra at the CCSD/def2-TZVPP level \cite{Pur82} with Gaussian16. \cite{Gaussian16} These calculations confirm the minima nature of the obtained 
geometries. \cite{zzz-tou-1} We then re-optimize the structures at the {\CC{3}}/{\AVTZ} level \cite{Chr95b,Koc97} using Dalton \cite{dalton} and/or CFOUR, \cite{cfour} depending on the size and symmetry of the molecule. 
CFOUR advantageously provides analytical CC3 gradients for ground-state structures. For the CCSD calculations, the energy and geometry convergence thresholds are  systematically tightened to \InAU{$10^{-10}$--$10^{-11}$}~for the 
SCF energy, \InAU{$10^{-8}$--$10^{-9}$}~for the {\CCSD} energy, and \InAU{$10^{-7}$--$10^{-8}$}~for the {\EOMCCSD} energy in the case of excited-state geometry optimizations.  To check that the structures correspond 
to genuine minima, the (EOM-){\CCSD} gradients are differentiated numerically to obtain the vibrational frequencies. The {\CC{3}} optimizations are performed with the default convergence thresholds of Dalton or CFOUR 
without applying the frozen core approximation. 

\subsection{Coupled Cluster calculations}

Unless otherwise stated, the {\CC{}} transition energies \cite{Kal04} are computed in the frozen-core approximation (large cores for \ce{Cl} and \ce{S}). We use several codes to achieve our objectives, namely CFOUR,\cite{cfour}  
Dalton,\cite{dalton} Gaussian16,\cite{Gaussian16} Orca,\cite{Nee12} MRCC,\cite{Rol13,mrcc}  and Q-Chem. \cite{Sha15} Globally, we use  CFOUR for both CCSDT-3 \cite{Wat96,Pro10} and CCSDT \cite{Nog87} calculations, 
Dalton to perform the CIS(D),\cite{Hea94,Hea95}  {\CC{2}}, \cite{Chr95,Hat00} {\CCSD},\cite{Pur82}  CCSDR(3), \cite{Chr96b} and CC3 \cite{Chr95b,Koc97} calculations, Gaussian for the CIS(D) \cite{Hea94,Hea95} and {\CCSD}, \cite{Pur82}  
Orca for  the similarity-transformed EOM-CCSD (STEOM-CCSD)\cite{Noo97,Dut18} calculations, Q-Chem  for  {\ADC{2}} and {\ADC{3}} calculations,  and MRCC for the CIS(D$_\infty$),\cite{Hea99}  {\CCSDT}, \cite{Nog87} CCSDTQ, \cite{Kuc91} (and higher) 
calculations. As we mainly report transition energies, it {is} worth noting that the linear-response (LR) and equation-of-motion (EOM) formalisms provide identical results. Nevertheless, the oscillator strengths characterizing the excited 
states are obtained at the (LR) {\CC{3}} level with Dalton. Default program setting are generally applied, and when modified they are tightened. For the STEOM-CCSD calculations which relies on natural transition orbitals, 
it was checked that each state is characterized by an active character percentage of 98\%\ or larger (states not matching this criterion are not reported). Nevertheless, the obtained results  do slightly depend on the number of 
states included in the calculations, and we found typical variations of $\pm$\IneV{0.01--0.05}. For all calculations, we use the well-known Dunning's \emph{aug}-cc-pVXZ (X $=$ D, T, Q and 5) atomic basis sets, as well as some 
doubly- and triply-augmented basis sets of the same series (d-\emph{aug}-cc-pVXZ and t-\emph{aug}-cc-pVXZ).

\subsection{Selected Configuration Interaction methods}

Alternatively to {\CC{}}, we also compute transition energies using a selected {\CI} ({\sCI}) approach, an idea that goes back to 1969 in the pioneering works of Bender and Davidson, \cite{Ben69} and Whitten and Hackmeyer. \cite{Whi69} 
Recently, sCI methods have demonstrated their ability to reach near FCI quality energies for small organic and transition metal-containing molecules. \cite{Gin13,Caf14,Gin15,Gar17,Caf16,Hol16,Sha17,Hol17,Chi18,Sce18}
To avoid the exponential increase of the size of the {\CI} expansion, we employ the {\sCI} algorithm CIPSI \cite{Hur73,Eva83,Gin13} (Configuration Interaction using a Perturbative Selection made Iteratively) to retain only the 
energetically-relevant determinants. To do so, the CIPSI algorithm uses a second-order energetic criterion to select perturbatively determinants in the {\FCI} space. \cite{Gin13,Gin15,Caf16,Sce18} 
In the numerical examples presented below, our CI expansions contain typically about a few millions of determinants. We refer the interested readers to Ref.~\citenum{Caf16b,Sce18} for more details about the general philosophy 
of {\sCI} methods. 

In order to treat the electronic states of a given spin manifold on equal footing, a common set of determinants is used for all states. Moreover, to speed up convergence to the {\FCI} limit, a common set of natural orbitals issued 
from a preliminary (smaller) {\sCI} calculation is employed.  
{All sCI calculations have been performed in the frozen-core approximation.}
For a given basis set, we estimate the {\FCI} limit using the approach introduced recently by Holmes \emph{et. al.}  \cite{Hol17}  in the context of the (selected) 
heat-bath {\CI} method, and used with success, even for challenging chemical situations.\cite{Sha17,Sce18,Chi18} More precisely, we linearly extrapolate the {\sCI} 
energy $\EsCI$ as a function of $\EPT$, which is an estimate of the truncation error in the {\sCI} algorithm, i.e., $\EPT \approx \EFCI-\EsCI$. When $\EPT = 0$, the {\FCI} limit has effectively been reached.  Here, $\EPT$ is 
efficiently evaluated with a recently-proposed hybrid stochastic-deterministic algorithm. \cite{Gar17b} Note that we do not report error bars because the statistical errors originating from this algorithm are orders of magnitude 
smaller than the extrapolation errors. In practice, the extrapolation is based on the two largest sCI wave functions, i.e., we perform a two-point extrapolation, which is justified here because of the quasi-linear
behavior of the {\sCI} energy as a function of $\EPT$. Estimating the extrapolation error is a complicated task with no 
well-defined method to do so. In practice, we have observed that this extrapolation procedure is robust and provides FCI estimates within \IneV{$\pm 0.02$}. When the convergence to the FCI limit is too slow to provide reliable 
estimates, the number of significant digits reported has been reduced accordingly. 
From herein, the extrapolated {\FCI} results are simply labeled {\exCI}. 
{Several illustrative examples are reported in Supporting Information where we compare different types of extrapolations for several molecules (See Fig.~S1 and Table S11).
In particular, diazomethane and streptocyanine-C1 can be considered as ``difficult'' cases (\textit{vide infra}), and the results reported in Supporting Information show that, even in these challenging situations, the two-point linear extrapolation is fairly robust.
Moreover, additional points do not significantly alter the exFCI results (typically 0.01 eV or less).}

All the {\sCI} calculations are performed with the electronic structure 
software \textsc{quantum package}, developed in Toulouse and freely available. \cite{QP} Additional information about the sCI wave functions, excitations energies as well as their extrapolated values can be found at the end
of the {\SI}.

%
%
\section{Results and Discussion}
\label{sec-res}

In the discussion below, we first discuss specific molecules of increasing size and compare the results obtained with {\exCI} and {\CC{}} approaches, starting with the {\CC{3}} method for the latter.
{This first part is performed applying systematically the frozen-core approximation.}
We next define two series of TBE, one at the frozen-core {\AVTZ} level, and one close to complete basis set limit by applying corrections for frozen-core and basis set effects. In the following stage, 
we assess the performances of several popular wave function methods using the former benchmark as reference. Finally, we discuss the performances of basis set extrapolation approaches
starting from a compact basis. Unless otherwise stated, we considered the {\exCI} values as benchmarks.

\subsection{Water, hydrogen sulfide, ammonia, and hydrogen chloride}

\begin{sidewaystable}[htp]
\caption{\small Vertical transition energies for the three lowest singlet and three lowest triplet excited states of water (top), the four lowest singlet and the lowest triplet states of ammonia (center), and the lowest singlet state of hydrogen chloride (bottom). 
All states of water and ammonia have a Rydberg character, whereas the lowest state of hydrogen chloride is a charge-transfer state. 
All values are in eV.} 
\label{Table-1}
  \begin{small}
\begin{tabular}{l|p{.6cm}p{1.1cm}p{1.4cm}p{1.7cm}p{.9cm}|p{.6cm}p{1.1cm}p{1.4cm}p{.9cm}|p{.6cm}p{1.1cm}p{.9cm}|p{.7cm}p{.7cm}p{.7cm}}
\hline 
		 \multicolumn{16}{c}{Water}\\
		& \multicolumn{5}{c}{\AVDZ} & \multicolumn{4}{c}{\AVTZ}& \multicolumn{3}{c}{\AVQZ} & \multicolumn{3}{c}{Litt.}\\
State 	& {\CC{3}} & {\CCSDT} & {\CCSDTQ} & {\CCSDTQP} & {\exCI} & {\CC{3}} & {\CCSDT} & {\CCSDTQ}  & {\exCI}& {\CC{3}} & {\CCSDT}   & {\exCI} & Exp.$^a$ & Th.$^b$ & Th.$^c$\\
\hline
$^1B_1 (n \rightarrow 3s)$ 	&7.51&7.50&7.53&7.53&7.53	&7.60&7.59&7.62&7.62	&7.65	&7.64	&7.68	&7.41 &7.81&7.57\\
$^1A_2 (n \rightarrow 3p)$ 	&9.29&9.28&9.31&9.32&9.32	&9.38&9.37&9.40&9.41	&9.43	&9.41	&9.46	&9.20 &9.30&9.33\\
$^1A_1 (n \rightarrow 3s)$ 	&9.92&9.90&9.94&9.94&9.94	&9.97&9.95&9.98&9.99	&10.00	&9.98	&10.02	&9.67 &9.91&9.91\\
$^3B_1 (n \rightarrow 3s)$ 	&7.13&7.11&7.14&7.14&7.14	&7.23&7.22&7.24&7.25	&7.28	&7.26	&7.30	&7.20 &7.42&7.21\\
$^3A_2 (n \rightarrow 3p)$ 	&9.12&9.11&9.14&9.14&9.14	&9.22&9.20&9.23&9.24	&9.26	&9.25	&9.28	&8.90 &9.42&9.19\\
$^3A_1 (n \rightarrow 3s)$ 	&9.47&9.45&9.48&9.49&9.49	&9.52&9.50&9.53&9.54	&9.56	&9.54	&9.58	&9.46 &9.78&9.50\\
\hline
		 \multicolumn{16}{c}{Hydrogen sulfide}\\
		& \multicolumn{5}{c}{\AVDZ} & \multicolumn{4}{c}{\AVTZ}& \multicolumn{3}{c}{\AVQZ} & \multicolumn{3}{c}{Litt.}\\
State 	& {\CC{3}} & {\CCSDT} & {\CCSDTQ} & {\CCSDTQP} & {\exCI} & {\CC{3}} & {\CCSDT} & {\CCSDTQ}  & {\exCI}& {\CC{3}} & {\CCSDT}   & {\exCI} & Exp.$^d$ & Exp.$^e$ & Th.$^f$\\
\hline
$^1A_2 (n \rightarrow 4p)$ 	&6.29&6.29&6.29&6.29&6.29	&6.19&6.18&6.18&6.18	&6.16&6.15&6.15	&	  &     &6.12\\
$^1B_1 (n \rightarrow 4s)$ 	&6.10&6.10&6.10&6.10&6.10	&6.24&6.24&6.24&6.24	&6.29&6.29&6.29 	&6.33 &     &6.27\\
$^3A_2 (n \rightarrow 4p)$ 	&5.91&5.90&5.90&5.90&5.90	&5.82&5.81&5.81&5.81	&5.80&5.79&5.79 	&	  &5.8&5.78\\
$^3B_1 (n \rightarrow 4s)$ 	&5.75&5.75&5.75&5.75&5.75	&5.88&5.88&5.88&5.89	&5.93&5.93&5.93 	&        &5.4&5.92 \\
\hline
		 \multicolumn{16}{c}{Ammonia}\\
		& \multicolumn{5}{c}{\AVDZ} & \multicolumn{4}{c}{\AVTZ}& \multicolumn{3}{c}{\AVQZ} & \multicolumn{3}{c}{Litt.}\\
State 	& {\CC{3}} & {\CCSDT} & {\CCSDTQ} & {\CCSDTQP} & {\exCI} & {\CC{3}} & {\CCSDT} & {\CCSDTQ}  & {\exCI}& {\CC{3}} & {\CCSDT}   & {\exCI} & Exp.$^g$ & Exp.$^h$ & Th.$^i$\\
\hline
$^1A_2 (n \rightarrow 3s)$ 	&6.46  &6.46  &	6.48 &6.48&6.48 	&6.57&6.57&6.59	&6.59	&6.61&6.61&6.64	&6.38&6.39	&6.48\\
$^1E (n \rightarrow 3p)$ 		&8.06  &8.06  &8.08 &8.08&8.08 	&8.15&8.14&8.16	&8.16	&8.18&8.17&8.22	&7.90&7.93	&8.02\\
$^1A_1 (n \rightarrow 3p)$ 	&9.66  &9.66  &9.68 &9.68&9.68 	&9.32&9.31&		&9.33	&9.11&9.10&9.14	&8.14&8.26	&8.50\\
$^1A_2 (n \rightarrow 4s)$ 	&10.40&10.39&10.41&10.41&10.41	&9.95&9.94&		&9.96	&9.77&9.77&		&	 &		&9.03\\
$^3A_2 (n \rightarrow 3s)$ 	&6.18  &6.18  &6.19  &6.19&6.19 	&6.29&6.29&6.30	&6.31	&6.33&6.33&6.35	&\emph{6.02}$^j$ &		&\\
\hline
		 \multicolumn{16}{c}{Hydrogen chloride}\\
		& \multicolumn{5}{c}{\AVDZ} & \multicolumn{4}{c}{\AVTZ}& \multicolumn{3}{c}{\AVQZ} & \multicolumn{1}{c}{Litt.}\\
State 	& {\CC{3}} & {\CCSDT} & {\CCSDTQ} & {\CCSDTQP} & {\exCI} & {\CC{3}} & {\CCSDT} & {\CCSDTQ}  & {\exCI}& {\CC{3}} & {\CCSDT}   & {\exCI} & Th.$^k$\\
\hline
$^1 \Pi (\mathrm{CT})$		&7.82&7.81&7.82&7.82		&7.82	&7.84&7.83&7.84	&7.84	&7.89&7.88$^l$ &7.88 &8.23\\
\hline
 \end{tabular}
  \end{small}
\begin{flushleft}
\begin{footnotesize}
$^a${Energy loss experiment from Ref.~\citenum{Ral13};}
$^b${MRCI+Q/{\AVTZ} calculations from Ref.~\citenum{Cai00c};}
$^c${MRCC/{\AVTZ} calculations from Ref.~\citenum{Li06b};}
$^d${VUV experiment from Ref.~\citenum{Mas79};}
$^e${Electron impact experiment from Ref.~\citenum{Abu86};}
$^f${{\CASPT}/{\DAVQZ} results from Ref.~\citenum{Pal08};}
$^g${Electron impact experiment from Ref.~\citenum{Ske65};}
$^h${Electron impact experiment from Ref.~\citenum{Har71};}
$^i${EOM-CCSDç($\tilde{T}$)/{\AVTZ} with extra \emph{diffuse} calculations from Ref.~\citenum{Bar97};}
$^j${Deduced from the \IneV{$6.38$} value of the $^1A_2 (n \rightarrow 3s)$ state and the \IneV{$-0.36$} shift reported for the 0-0 energies compared 
to the corresponding singlet state in Ref.~\citenum{Ben91}, a splitting consistent with an earlier estimate of \IneV{$-0.39$} given in Ref.~\citenum{Abu84};}
$^k${CC2/cc-pVTZ from Ref.~\citenum{Pea08};}
$^l${The {\CCSDTQ}/{\AVQZ} value is \IneV{7.88} as well.}
\end{footnotesize}
\end{flushleft}
\end{sidewaystable}

Due to its small size and ubiquitous role in life, water is often used as a test case for Rydberg excitations. Indeed, it is part of Head-Gordon's \cite{Rhe07}, Gordon's \cite{Lea12} and Truhlar-{Gagliardi}'s \cite{Hoy16} datasets of 
compounds, and it has been investigated at many levels of theory. \cite{Cai00c,Li06b,Rub08,Pal08}  Our results are collected in Table \ref{Table-1}. With the {\AVDZ} basis, there is an nearly perfect agreement between the 
{\exCI} values and the transition energies obtained with the two largest {\CC{}} expansions, namely {\CCSDTQ} and {\CCSDTQP}. Indeed, the largest discrepancy is as small as \IneV{$0.01$}, and it is therefore reasonable 
to state that the {\FCI} limit has been reached with that specific basis set. Compared to the {\exCI} results, the {\CCSDT} values are systematically too low, with an average error of \IneV{$-0.03$}. The same trend of underestimation 
is found  with {\CC{3}}, though with smaller absolute deviations for all states.  Unsurprisingly, for Rydberg states, increasing the basis set size has a significant impact, and it tends to increase the computed transition energies in 
water. However, this effect is very similar for all methods listed in Table \ref{Table-1}. This means that, on the one hand,  the tendency of  {\CCSDT}  to provide slightly too small transition energies pertains with both {\AVTZ} and 
{\AVQZ}, and, on the other hand, that estimating the basis set effect with a ``cheap'' method is possible. Indeed, adding to the {\exCI}/{\AVDZ} energies, the difference between {\CC{3}}/{\AVQZ} and {\CC{3}}/{\AVDZ} results would 
deliver estimates systematically within \IneV{$0.01$} of the actual {\exCI}/{\AVQZ} values. Such basis set extrapolation approach was already advocated for lower-order {\CC{}} expansions, \cite{Sil10b,Jac15a} and it is therefore 
not surprising that it can be applied with refined models.  As it can be seen in Table S1 in the {\SI}, further extension of the basis set or correlation of the $1s$ electron have small impacts, except for the Rydberg $^1A_1$ state. 
Eventually, as evidenced by the data from the rightmost columns of Table \ref{Table-1}, the present estimates are in good agreement with previous MRCC values determined on the experimental geometry, \cite{Li06b} whereas 
the experimental values offer qualitative comparisons only, for reasons discussed elsewhere. \cite{Ral13} We underline that some of the 2013 measurements reported in Table \ref{Table-1} significantly differ from previous electron 
impact data, \cite{Chu75} that were used previously as reference, \cite{Lea12} with e.g., a \IneV{$0.2$} discrepancy between the two experiments for the lowest triplet state.

As water, hydrogen sulfide was also the subject of several high-level theoretical investigation, \cite{Pit02,Vel04,Gup07,Pal08} which are necessary, as  there are rather few experimental data available for the lowest Rydberg states
of H$_2$S, \cite{Mas79,Obr83,Rob85b,Abu86} especially no accurate value could be measured for the first $^1A_2$ state.  As can be seen in Table \ref{Table-1}, for a given basis set all tested {\CC{}} methods provide very similar 
results, systematically within \IneV{0.01} of the {\exCI} results. In contrast, the basis set has a significant impact, e.g., the two lowest singlet states switch order when going from {\AVDZ} to {\AVTZ} and the same is true for the two 
lowest triplet states.  Our results are also very consistent with the  {\CASPT}/{\DAVQZ} values given in Ref.~\citenum{Pal08}, confirming that a near {\FCI} limit has been reached.

Ammonia is also another molecule for evaluating Rydberg excitations, and it was previously investigated at several levels of theory. \cite{Cha91b,Bar97,Rhe07,Sch17} As in the case of water, we note a nearly perfect match 
between the  {\CCSDTQ} and  {\exCI} estimates with both the {\AVDZ} and {\AVTZ} atomic basis sets, indicating that the {\FCI} limit is reached. Both {\CC{3}} and {\CCSDT} are close to this limit, and the former model slightly 
outperforms the latter. For ammonia, the basis set effects are particularly strong for the third and fourth singlet excited states but these basis set effects are nearly transferrable from one method to another. In fact, as hinted 
by the large differences between the {\AVTZ} and {\AVQZ} results in Table \ref{Table-1}, these two high-lying states require the use of additional diffuse orbitals to attain convergence. The {\CC{3}}/{\TAVQZ} values of $8.60$ and 
\IneV{$9.15$} (see Table S1 in the {\SI}), are close from the previous results of Bartlett and coworkers, \cite{Bar97} who also applied extra diffuse orbitals in their calculations relying on approximate triples (see the footnotes in 
Table \ref{Table-1}).  As in water, the experimental values do not provide sufficiently clear-cut results to ultimately decide which method is the most accurate. {Indeed, the vertical experimental estimates reported in Table} \ref{Table-1}
{differ significantly from the more trustworthy adiabatic values with variations of ca. 0.5 eV}.\cite{Bar97} {Consequently, a good match between an experimental measurement and a theoretical calculation determined with
a compact basis set is, in the present case, a sign of lucky cancellation of errors.}

Hydrogen chloride was less frequently used in previous benchmarks, but is included in Tozer's set as an example of charge-transfer (CT) state. \cite{Pea08} Again, the results listed at the bottom of Table \ref{Table-1} 
demonstrate a remarkable consistency between the various theories. Though large frozen cores are used during the calculations, this does not strongly impact the results, as can be deduced from the data of Table S1.  
As expected, the absorption band corresponding to this CT state is very broad experimentally (starting at \IneV{$5.5$} and peaking at \IneV{$8.1$}), \cite{Hub79} making direct comparisons tricky.

\subsection{Dinitrogen and carbon monoxide}

\begin{sidewaystable}[htp]
\caption{\small Vertical transition energies for various excited states of dinitrogen (top) and carbon monoxide (bottom).  R stands for Rydberg state. All values are in eV.} 
\label{Table-2}
  \begin{small}
\begin{tabular}{l|p{.6cm}p{1.1cm}p{1.4cm}p{1.7cm}p{.9cm}|p{.6cm}p{1.1cm}p{1.4cm}p{.9cm}|p{.6cm}p{1.1cm}p{.9cm}|p{.7cm}p{.7cm}p{.7cm}}
\hline 
		 \multicolumn{16}{c}{Dinitrogen}\\
		& \multicolumn{5}{c}{\AVDZ} & \multicolumn{4}{c}{\AVTZ}& \multicolumn{3}{c}{\AVQZ} & \multicolumn{3}{c}{Litt.}\\
State 	& {\CC{3}} & {\CCSDT} & {\CCSDTQ} & {\CCSDTQP} & {\exCI} & {\CC{3}} & {\CCSDT} & {\CCSDTQ}  & {\exCI}& {\CC{3}} & {\CCSDT}   & {\exCI} & Exp.$^a$ &  Exp.$^b$ & Th.$^c$\\
\hline								
$^1\Pi_g (n \rightarrow \pis)$ 				&9.44	&9.41  &	9.41	&9.41 &	9.41		&9.34 &9.33 	&9.32	&9.34	&9.33 	&9.31	&9.34	&9.31	&9.31	&9.27	\\
$^1\Sigma_u^- (\pi \rightarrow \pis)$			&10.06	&10.06&	10.06&10.05&	10.05	&9.88 &9.89 	&9.88	&9.88	&9.87 	&9.88	&9.92	&9.92	&9.92	&10.09	\\
$^1\Delta_u (\pi \rightarrow \pis)$ 			&10.43	&10.44&	10.43&10.43&	10.43	&10.29&10.30 	&		&10.29	&10.27 	&10.28	&10.31	&10.27	&10.27	&10.54	\\
$^1\Sigma_g^+ (\Ryd)$ 					&13.23	&13.20&	13.18&13.18&	13.18	&13.01&13.00	&12.97	&12.98	&12.90 	&12.89	&12.89	&		&12.2	&12.20	\\
$^1\Pi_u (\Ryd)$ 						&13.28	&13.17&	13.13&13.13&	13.12	&13.22&13.14	&13.09	&13.03	&13.17 	&		&13.1$^d$&12.78	&12.90	&12.84	\\
$^1\Sigma_u^+ (\Ryd)$ 					&13.14	&13.13&	13.11&13.11&	13.11	&13.12&13.12	&13.09	&13.09	&13.09 	&13.09	&13.2$^d$&12.96	&12.98	&12.82	\\
$^1\Pi_u (\Ryd)$ 						&13.64	&13.59&	13.56&13.56&	13.56	&13.49&13.45	&13.42	&13.46	&13.42 	&13.37	&13.7$^d$&13.10	&13.24	&13.61	\\
$^3\Sigma_u^+ (\pi \rightarrow \pis)$			&7.67	&7.68&	7.69	&7.70 &	7.70		&7.68 &7.69	&7.70	&7.70	&7.71 	&7.71	&7.74	&7.75	&7.75	&7.56	\\
$^3\Pi_g (n \rightarrow \pis)$ 				&8.07	&8.06&	8.05 &8.05 &	8.05		&8.04 &8.03	&8.02	&8.01	&8.04 	&8.04	&8.03	&8.04	&8.04	&8.05	\\
$^3\Delta_u (\pi \rightarrow \pis)$ 			&8.97	&8.96&	8.96 &8.96 &	8.96		&8.87 &8.87	&8.87	&8.87	&8.87 	&8.87	&8.88	&8.88	&8.88	&8.93	\\
$^3\Sigma_u^- (\pi \rightarrow \pis)$			&9.78	&9.76&	9.75 &9.75&	9.75		&9.68 &9.68	&9.66	&9.66	&9.68 	&		&9.66	&9.67	&9.67	&9.86	\\
\hline
		 \multicolumn{16}{c}{Carbon monoxide}\\
		& \multicolumn{5}{c}{\AVDZ} & \multicolumn{4}{c}{\AVTZ}& \multicolumn{3}{c}{\AVQZ} & \multicolumn{3}{c}{Litt.}\\
State 	& {\CC{3}} & {\CCSDT} & {\CCSDTQ} & {\CCSDTQP} & {\exCI} & {\CC{3}} & {\CCSDT} & {\CCSDTQ}  & {\exCI}& {\CC{3}} & {\CCSDT}   & {\exCI} & Exp.$^e$ & Th.$^f$ & Th.$^g$\\
\hline								
$^1\Pi (n \rightarrow \pis)$ 				&8.57  &8.57  &8.56	&8.56&   8.57	&8.49  &8.49  &8.48&  8.49	&8.47 &8.48  &  8.50		&8.51	&8.54	&8.83	\\
$^1\Sigma^- (\pi \rightarrow \pis)$			&10.12&10.06&10.06&10.06&10.05	&9.99  &9.94  &	9.93&  9.92	&9.99 &9.94  &  9.99		&9.88	&10.05	&9.97	\\
$^1\Delta (\pi \rightarrow \pis)$ 				&10.23&10.18&10.17&10.17&10.16	&10.12&10.08&	10.07& 10.06	&10.12&10.07& 10.11	&10.23	&10.18	&10.00	\\
$^1\Sigma^+ (\Ryd)$ 					&10.92&10.94&10.93&10.92& 10.94	&10.94&10.99&	10.96&10.95	&10.90&10.95& 10.96	&10.78	&10.98	&		\\
$^1\Sigma^+ (\Ryd)$ 					&11.48&11.52&11.51	&11.51&  11.52	&11.49&11.54&11.52	& 11.52	&11.46&11.51&  11.53	&11.40	&		&		\\
$^1\Pi (\Ryd)$							&11.74&11.77&11.76	&11.75&  11.76	&11.69&11.74&	11.72& 11.72	&11.63&11.69&  11.70	&11.53	&		&		\\
$^3\Pi (n \rightarrow \pis)$ 				&6.31  &6.30  &6.29	&6.28&    6.29	&6.30  &6.30  &6.28	& 6.28	&6.30  &6.30  &	  6.29	&6.32	&		&6.41	\\
$^3\Sigma^+ (\pi \rightarrow \pis)$			&8.45  &8.43  &8.44	&8.44&     8.46	&8.45  &8.42  &	8.44& 8.45	&8.48  &8.45  &	  8.49	&8.51	&		&8.39	\\
$^3\Delta (\pi \rightarrow \pis)$ 				&9.37  &9.33  &9.34	&9.34&    9.33	&9.30  &9.26  &	9.26&  9.27	&9.31  &9.26  &	  9.29	&9.36	&		&9.23	\\
$^3\Sigma^- (\pi \rightarrow \pis)$			&9.89  &	      &		&	&    9.83	&9.82  &	      &		& 9.80	&9.82  &	      &  9.78	&9.88	&		&9.60	\\
$^3\Sigma^+ (\Ryd)$ 					&10.39&10.42&10.42&10.41&10.41	&10.45&10.50&10.48& 10.47	&10.44&10.49&			&10.4$^h$&		&		\\
\hline
 \end{tabular}
  \end{small}
\begin{flushleft}
\begin{footnotesize}
$^a${Experimental vertical values given in Ref.~\citenum{Odd85} and computed from the spectroscopic constants of Ref.~\citenum{Hub79};}
$^b${Experimental vertical values given in Ref.~\citenum{Ben90} and computed from the spectroscopic constants of Ref.~\citenum{Hub79};}
$^c${MRCCSD/6-311G with one additional $d$ calculations from Ref.~\citenum{Ben90};}
$^d${{\CI} convergence too slow to provide reliable estimates};
$^e${Experimental vertical values given in Ref.~\citenum{Nie80b} and computed from the spectroscopic constants of Ref.~\citenum{Hub79};}
$^f${CCSDT/PVTZ+ results from Ref.~\citenum{Kuc01};}
$^g${CASSCF(10,10)/cc-pVTZ results from Ref.~\citenum{Dor16};}
$^h${Only one digit reported for that state, see Ref.~\citenum{Nie80b}.}
\end{footnotesize}
\end{flushleft}
\end{sidewaystable}
		
Dinitrogen is a simple diatomic compound for which the low-lying valence and Rydberg states have been investigated at several levels of theory. \cite{Odd85,Ben90,Kuc01,Pea08} With a numerical solution of the nuclear Schr{\"o}dinger equation, 
it is possible to treat the experimental spectroscopic constants, \cite{Hub79} so as to obtain reliable vertical estimates, and this procedure was applied previously. \cite{Sta83,Odd85,Ben90} Whilst such approach is supposedly providing experimental 
vertical excited-state energies with a ca.~\IneV{0.01} error only, it remains that significant excitation energy differences have been reported for the two lowest $^1\Pi_u$ states (see Table \ref{Table-2}). As in the previous cases, we find a remarkable 
agreement between the  {\CCSDTQ} and  {\exCI} estimates for most cases in which both could be determined. The only exceptions are the two $^1\Pi_u$ states with the {\AVTZ} basis, but in these two cases, the {\CC{}} expansion is also 
converging more slowly than usual, which is consistent with the relatively small degree of single excitation character in these two states (82.9 and 87.4\% according to {\CC{3}}). In contrast to water and ammonia, {\CCSDT} outperforms {\CC{3}} 
with respective mean absolute deviation (MAD) compared to {\exCI} of \IneV{$0.02$} and \IneV{$0.04$}, when using the {\AVDZ} basis set. As it can be deduced from Table S2 in the {\SI}, the basis set corrections are negligible for all valence states, 
but significant for some of the Rydberg states, especially, $^1\Sigma_g^+$ that requires two sets of diffuse orbitals to be reasonably close from the basis set limit. Applying  {\CC{3}}/{\DAVPZ} corrections to the most accurate {\exCI}  data, once 
can determine TBE values (\emph{vide infra}) that deviate only by \IneV{$0.02$} on (absolute) average compared to the experimental estimates for the seven valence states of dinitrogen.  Considering the expected inaccuracy of  \IneV{$0.01$} 
of the reference values, chemical accuracy is obviously reached without any experimental input. The deviations are about twice larger for the Rydberg states. Nevertheless, for the two $^1\Pi_u$ states, our TBE values, determined on the basis 
of {\exCI}/{\AVTZ} are  \IneV{$12.73$} and  \IneV{$13.27$} (\emph{vide infra}). This indicates  that for the lowest $^1\Pi_u$ state the estimate of Ref.~\citenum{Odd85} (\IneV{$12.78$}) is probably more accurate than the one of 
Ref.~\citenum{Ben90}  (\IneV{$12.90$}), whereas the opposite is likely true for the highest  $^1\Pi_u$ state that was reported to be located at \IneV{$13.10$} and \IneV{$13.24$} in Refs.~\citenum{Odd85} and  \citenum{Ben90}, respectively.  
One could argue that reaching agreement between CI and CC is particularly challenging for these two states. However, performing the basis set extrapolation starting from the {\CCSDTQP}/{\AVDZ} results would yield similar TBE of \IneV{12.77} and \IneV{$13.22$}.

For the isoelectronic carbon monoxide, experimental vertical energies deduced from rovibronic data\cite{Hub79} using a numerical approach are also available. \cite{Nie80b,Pea08} With the {\AVTZ} ({\AVQZ}) atomic basis set, the {\CCSDT} 
and {\CC{3}} results are within \IneV{$0.02$} (\IneV{$0.03$}) and \IneV{$0.03$} (\IneV{$0.03$}) of the {\exCI} results, whereas the errors made by both {\CCSDTQ} and {\CCSDTQP} are again trifling. As for dinitrogen, all the valence states are 
rather close from the basis set limit with {\AVTZ}, whereas larger basis sets are required for the Rydberg states (Table S2). By correcting the {\exCI}/{\AVQZ} ({\exCI}/{\AVTZ} for the highest triplet state) data with basis set effects determined 
at the {\CC{3}}/{\DAVPZ} level, we obtain TBE values that can be compared to the experimental estimates. The computed MAD is \IneV{$0.05$}, the largest deviations being obtained for the $\Delta$ and $\Sigma^-$ excited states of both 
spin symmetries.  The agreement between theory and experiment is therefore very satisfying though slightly less impressive than for \ce{N2}. We note that the {\CC{3}}/{\AVTZ} \ce{C=O} bond length (\InAA{$1.134$}) is \InAA{0.006} larger than the 
experimental $r_e$ value of \InAA{1.128},\cite{Hub79} whereas the discrepancy is twice smaller for dinitrogen: \InAA{$1.101$} for {\CC{3}}/{\AVTZ} compared to \InAA{$1.098$} experimentally. This might partially explained the larger deviations 
noticed for carbon monoxide.
	
\subsection{Acetylene and ethylene}

Acetylene is the smallest conjugated organic molecule possessing stable low-lying excited-state structures, therefore allowing to investigate vertical fluorescence.  This molecule has been the subject of previous investigations at the 
{\CASPT},\cite{Mal98} {\CCSD},\cite{Zyu03} {\CCSDT},\cite{Kan17} and MR-AQCC\cite{Ven03} levels. Our results are collected in Table \ref{Table-3}. With the double-$\zeta$ basis set, the differences between the {\CC{3}}, {\CCSDT}, 
and {\CCSDTQ} results are negligible, and the latter estimates are also systematically within \IneV{$0.02$} of the {\exCI} results. In contrast to water and ammonia, both {\CC{3}} and {\CCSDT} provide similar accuracies  compared 
to higher levels of theory.  As expected, for valence states, going from double- to triple-$\zeta$ basis set tends to slightly decrease the computed energies (except for the lowest triplet). Nonetheless, as with the smaller basis set, the same near-perfect 
methodological match pertains with {\AVTZ}. Estimating the {\exCI}/{\AVTZ} results from the {\exCI}/{\AVDZ} values and {\CC{3}} basis set effects would yield estimates with absolute errors of \IneV{$0.00$--$0.02$}. One also notices that 
the {\exCI}/{\AVTZ}  values are all extremely close to the previous MR-AQCC estimates, whereas the published {\CASPT} values appear to be too low {though closer from the electron impact experiment, underlying once 
more the difficulty to obtain very accurate experimental estimates for vertical energies.} This underestimating trend of standard {\CASPT} was reported before for other molecules.\cite{Ang05b,Sen11} Although our theoretical 
vertical energy estimates still slightly vary when passing from the {\AVDZ} to  {\AVTZ} basis sets, we claim that these vertical energies are  probably more trustworthy for further benchmarks than the available experimental values 
because basis set effects beyond  {\AVTZ} seem rather limited (Table S3).

\begin{table}[htp]
\caption{\small Vertical (absorption) transition energies for the five lowest low-lying valence excited states of acetylene (top) and the three lowest singlet and triplet
excited states of ethylene (bottom).  For acetylene, we also report the vertical emission (denoted [F]) obtained from the lowest \emph{trans} and \emph{cis} 
isomers. All values are in eV.}
\label{Table-3}
  \begin{small}
\begin{tabular}{l|cccc|ccc|ccc}
\hline 
		 \multicolumn{11}{c}{Acetylene}\\
		& \multicolumn{4}{c}{\AVDZ} & \multicolumn{3}{c}{\AVTZ} & \multicolumn{3}{c}{Litt.}\\
State 	& {\CC{3}} & {\CCSDT} & {\CCSDTQ} &  {\exCI} & {\CC{3}} & {\CCSDT}  & {\exCI}& Exp.$^a$ & Th.$^b$ & Th.$^c$ \\
\hline
$^1\Sigma_u^- (\pi \rightarrow \pis)$ 			&7.21&7.21&7.21&7.20	&7.09&7.09&7.10	&7.1&6.96&7.10\\
$^1\Delta_u	(\pi \rightarrow \pis)$ 		&7.51&7.52&7.52&7.51	&7.42&7.43&7.44	&7.2&7.30&7.43\\
$^3\Sigma_u^+ (\pi \rightarrow \pis)$ 		&5.48&5.49&5.50&5.50	&5.50&5.51&5.53	&5.2&5.26&5.58\\
$^3\Delta_u	(\pi \rightarrow \pis)$ 		&6.46&6.46&6.46&6.46	&6.40&6.39&6.40	&6.0&6.20&6.41\\
$^3\Sigma_u^- (\pi \rightarrow \pis)$ 			&7.13&7.14&7.14&7.14	&7.07&       &7.08	&7.1&6.90&7.05\\
$^1A_u [\mathrm{F}]	(\pi \rightarrow \pis)$		&3.70&3.72&3.70&3.71	&3.64&3.66&3.64	&&\\
$^1A_2 [\mathrm{F}]	(\pi \rightarrow \pis)$		&3.92&3.94&3.93&3.93	&3.84&3.86&3.85	&&\\
\hline
		 \multicolumn{11}{c}{Ethylene}\\
		& \multicolumn{4}{c}{\AVDZ} & \multicolumn{3}{c}{\AVTZ} & \multicolumn{2}{c}{Litt.}\\
State 	& {\CC{3}} & {\CCSDT} & {\CCSDTQ} &  {\exCI} & {\CC{3}} & {\CCSDT}  & {\exCI}& Exp.$^d$ & Th.$^e$  \\
\hline
$^1B_{3u} (\pi \rightarrow 3s)$ 		&7.29&7.29&7.30&7.31	&7.35&7.37&7.39	&7.11	&7.45\\
$^1B_{1u} (\pi \rightarrow \pis)$ 	&7.94&7.94&7.93&7.93	&7.91&7.92&7.93	&7.60	&8.00\\
$^1B_{1g} (\pi \rightarrow 3p)$ 		&7.97&7.98&7.99&8.00	&8.03&8.04&8.08	&7.80	&8.06\\
$^3B_{1u} (\pi \rightarrow \pis)$ 	&4.53&4.54&4.54&4.55	&4.53&4.53&4.54	&4.36	&4.55\\
$^3B_{3u} (\pi \rightarrow 3s)$ 		&7.17&7.18&7.18&7.16	&7.24&7.25&$^f$	&6.98	&7.29\\
$^3B_{1g} (\pi \rightarrow 3p)$ 		&7.93&7.94&7.94&7.93	&7.98&7.99&$^f$	&7.79	&8.02\\
\hline
 \end{tabular}
  \end{small}
\begin{flushleft}
\begin{footnotesize}
$^a${Electron impact experiment from Ref.~\citenum{Dre87}. Note that the \IneV{7.1} value for the $\Sigma_u^-$ singlet and triplet states should be viewed as a tentative assignment;}
$^b${LS-CASPT2/\emph{aug}-ANO calculations from Ref.~\citenum{Mal98};}
$^c${MR-AQCC/{extrap.}~calculations from Ref.~\citenum{Ven03};}
$^d${Experimental values collected from various sources from Ref.~\citenum{Rob85b} (see discussions in Refs.~\citenum{Ser93,Sch08} and  \citenum{Fel14});}
$^e${Best composite theory from Ref.~\citenum{Fel14}, close to {\FCI};}
$^f${{\CI} convergence too slow to provide reliable estimates.}
\end{footnotesize}
\end{flushleft}
\end{table}

Despite its small size, ethylene remains a challenging molecule and is included in many benchmark sets. \cite{Hea94,Sch08,She09b,Car10,Lea12,Hoy16} The assignments of the experimental data has been the subject of countless works, 
and we refer the interested readers to the discussions in Refs.~\citenum{Rob85b,Ser93,Sch08,Ang08,Fel14,Chi18}. On the theoretical side, the most complete and accurate investigation dedicated to the excited states of ethylene 
is due to Davidson's group who performed refined {\CI} calculations. \cite{Fel14}  They indeed obtained highly-accurate transition energies for ethylene, including for the valence yet challenging $^1B_{1u}$ state.  From our data, collected 
in Table \ref{Table-3}, one notices that the differences between {\exCI}/{\AVDZ} and {\CCSDTQ}/{\AVDZ} results are again trifling, the largest deviation being obtained for the $^3B_{3u} (\pi \rightarrow 3s)$ Rydberg state (\IneV{$0.02$}). 
In addition, given the nice agreement between {\CC{3}}, {\CCSDT} and {\exCI} values, one can directly compare  our {\CC{3}}/{\AVPZ} results (Table S3) to the values of reported in Ref.~\citenum{Fel14}: a mean absolute 
deviation (MAD) of \IneV{$0.03$} is obtained. The fact that our transition energies tend to be slightly smaller than Davidson's is likely due to geometrical effects. Indeed, our {\CC{3}}/{\AVTZ} \ce{C=C} distance is \InAA{1.3338}, i.e.,
 slightly longer than the best estimate provided in Davidson's work (\InAA{1.3305}).  Recently, a stochastic heat-bath {\CI} (SHCI)/ANO-L-pVTZ work reported \IneV{$4.59$} and \IneV{$8.05$} values for the $^3B_{1u}$ and $^1B_{1u}$ states, 
 respectively, \cite{Chi18} and we also ascribe the differences with our results to the use of a MP2 geometry in Ref.~\citenum{Chi18}. Interestingly, these authors found quite large discrepancies between their SHCI and their {\CC{}} results. 
Indeed, they reported CR-EOMCC(2,3)D estimates significantly larger than their SHCI results with \IneV{$+0.17$} and \IneV{$+0.20$}  upshifts for the triplet and singlet states, respectively. 
This highlights that only high-level {\CC{}} schemes are able to recover the  {\exCI} (or SHCI) results for ethylene.

\subsection{Formaldehyde, methanimine and thioformaldehyde}

\begin{table}[htp]
\caption{\small Vertical (absorption) transition energies for various excited states of formaldehyde (top), methanimine (center), and thioformaldehyde (bottom). 
 All values are in eV.}
\label{Table-4}
  \begin{small}
\begin{tabular}{l|cccc|ccc|ccc}
\hline 
		 \multicolumn{11}{c}{Formaldehyde}\\
		& \multicolumn{4}{c}{\AVDZ} & \multicolumn{3}{c}{\AVTZ} & \multicolumn{3}{c}{Litt.}\\
State 	& {\CC{3}} & {\CCSDT} & {\CCSDTQ} &  {\exCI} & {\CC{3}} & {\CCSDT}  & {\exCI}& Exp.$^a$ & Th.$^b$ & Th.$^c$ \\
\hline
$^1A_2 (n \rightarrow \pis)$ 				&4.00&3.99&4.00&3.99	&3.97&3.95&3.98	&4.07	&3.98	&3.88 \\
$^1B_2 (n \rightarrow 3s)$ 				&7.05&7.04&7.09&7.11	&7.18&7.16&7.23	&7.11 	&7.12	&\\
$^1B_2 (n \rightarrow 3p)$ 				&8.02&8.00&8.04&8.04	&8.07&8.07&8.13	&7.97 	&7.94	&8.11\\
$^1A_1 (n \rightarrow 3p)$ 				&8.08&8.07&8.12&8.12	&8.18&8.16&8.23	&8.14 	&8.16	&\\
$^1A_2 (n \rightarrow 3p)$ 				&8.65&8.63&8.68&8.65	&8.64&8.61&8.67	&8.37 	&8.38	&\\
$^1B_1 (\sigma \rightarrow \pis)$			&9.31&9.29&9.30&9.29	&9.19&9.17&9.22	& 		&9.32	&9.04\\
$^1A_1 (\pi \rightarrow \pis)$				&9.59&9.59&9.54&9.53	&9.48&9.49&9.43	& 		&9.83	&9.29\\
$^3A_2 (n \rightarrow \pis)$ 				&3.58&3.57&3.58&3.58	&3.57&3.56&3.58	&3.50	&		&3.50\\
$^3A_1 (\pi \rightarrow \pis)$				&6.09&6.08&6.09&6.10	&6.05&6.05&6.06	&5.86	&		&5.87\\
$^3B_2 (n \rightarrow 3s)$ 				&6.91&6.90&6.95&6.95	&7.03&7.02&7.06	&6.83	&		&\\
$^3B_2 (n \rightarrow 3p)$ 				&7.84&7.82&7.86&7.87	&7.92&7.90&7.94	&7.79	&		&\\
$^3A_1 (n \rightarrow 3p)$ 				&7.97&7.95&8.00&8.01	&8.08&8.06&8.10	&7.96	&		&\\
$^3B_1 (n \rightarrow 3d)$ 				&8.48&8.47&8.48&8.48	&8.41&8.40&8.42	& 		&		&\\
$^1A'' [\mathrm{F}]	(n \rightarrow \pis)$		&2.87&2.84&2.86&2.86	&2.84&2.82&2.80	& 		&		&\\
\hline
		 \multicolumn{11}{c}{Methanimine}\\
		& \multicolumn{4}{c}{\AVDZ} & \multicolumn{3}{c}{\AVTZ} & \multicolumn{3}{c}{Litt.}\\
State 	& {\CC{3}} & {\CCSDT} & {\CCSDTQ} &  {\exCI} & {\CC{3}} & {\CCSDT}  & {\exCI}& Th.$^d$ & Th.$^e$ \\
\hline
$^1A''(n \rightarrow \pis)$ 				&5.26&5.24&5.25&5.25	&5.20&5.19&5.23	&		5.32 & 5.18 \\
$^3A'' (n \rightarrow \pis)$ 			&4.63&4.63&4.63&4.63	&4.61&4.61&4.65	&		\\
\hline
		 \multicolumn{11}{c}{Thioformaldehyde}\\
		& \multicolumn{4}{c}{\AVDZ} & \multicolumn{3}{c}{\AVTZ} & \multicolumn{2}{c}{Litt.}\\
State 	& {\CC{3}} & {\CCSDT} & {\CCSDTQ} &  {\exCI} & {\CC{3}} & {\CCSDT}  & {\exCI}& Exp.$^a$ & Exp.$^f$  \\
\hline
$^1A_2 (n \rightarrow \pis)$ 				&2.27&2.25&2.26&2.26		&2.23&2.21&2.22	& 		&2.03	&\\
$^1B_2 (n \rightarrow 4s)$ 				&5.80&5.80&5.82&5.83		&5.91&5.89&5.96	&5.85	&5.84	&\\
$^1A_1 (\pi \rightarrow \pis)$				&6.62&6.60&6.51&6.5$^g$	&6.48&6.47&6.4$^g$&6.2		&5.54	&\\
$^3A_2 (n \rightarrow \pis)$ 				&1.97&1.96&1.96&1.97		&1.94&1.93&1.94	&		&1.80	&\\
$^3A_1 (\pi \rightarrow \pis)$				&3.43&3.43&3.44&3.45		&3.38&3.38&3.43	&3.28	&		&\\	
$^3B_2 (n \rightarrow 4s)$ 				&5.64&5.63&5.65&5.66		&5.72&5.71&5.6$^g$	&		&		&\\
$^1A_2 [\mathrm{F}]	(n \rightarrow \pis)$		&2.00&2.00&1.98&1.98		&1.97&1.98&1.95	&		&		&\\
\hline
 \end{tabular}
  \end{small}
\begin{flushleft}
\begin{footnotesize}
$^a${Various experimental sources, summarized in Ref.~\citenum{Rob85b};}
$^b${MR-AQCC-LRT calculations from Ref.~\citenum{Mul01};}
$^c${{\CC{3}}/{\AVQZ} calculations from Ref.~\citenum{Sch08};}
$^d${DMC results form Ref.~\citenum{Sch04e};}
$^e${{\CCSDT}/{\AVTZ} calculations from Ref.~\citenum{Kan17};}
$^f${0-0 energies collected in Ref.~\citenum{Pao84};}
$^g${{\CI} convergence too slow to provide reliable estimates.}
\end{footnotesize}
\end{flushleft}
\end{table}

Similarly to ethylene, formaldehyde is a very popular test molecule, \cite{For92b,Had93,Hea94,Hea95,Gwa95,Wib98,Wib02,Pea08,Sch08,She09b,Car10,Li11,Lea12,Hoy16,Kan17} and stands as the prototype carbonyl dye with a low-lying 
$n \rightarrow \pis$ transition.  Nevertheless, even for this particular valence state, well-separated from higher-lying excited states, the choice of an experimental reference remains difficult. Indeed, values of \IneV{$3.94$},\cite{Pea08} 
\IneV{$4.00$}, \cite{Had93,Car10,Hoy16} \IneV{$4.07$}, \cite{Hea94,Gwa95,Lea12} and \IneV{$4.1$}, \cite{For92b,Wib98} have been used in previous theoretical benchmarks. In contrast to their oxygen cousin, both methanimine and 
thioformaldehyde were the subject of less attention from the theoretical community. \cite{Fur02,Sch04e,Hat05c} The results obtained for these three molecules are collected in Table \ref{Table-4}.  Considering all transitions listed in this Table, one 
obtains a MAD of \IneV{$0.01$} between the {\CCSDTQ}/{\AVDZ} and {\exCI}/{\AVDZ} results, the largest discrepancies of \IneV{$0.03$} being observed for two states for which the difference between {\CCSDT} and {\CCSDTQ} 
is also large (\IneV{$0.05$}). As in water, using the {\exCI}/{\AVDZ} values as reference, we found that {\CC{3}} delivers slightly more accurate transition energies (MAD of \IneV{$0.02$}, maximal deviation of \IneV{0.06}) 
than {\CCSDT} (MAD of \IneV{0.03}, maximal deviation of \IneV{0.07}). By adding the difference between {\CC{3}}/{\AVTZ} and {\CC{3}}/{\AVDZ} results to the {\exCI}/{\AVDZ} values, we obtain good estimates of the actual 
{\exCI}/{\AVTZ} data, with a MAD of \IneV{$0.02$} for formaldehyde.  Compared to the {\CC{3}}/{\AVQZ} results of Thiel, \cite{Sch08} the transition energies reported in Table \ref{Table-4} are slightly larger, which is probably 
due to the influence of the ground-state geometry rather than basis set effects (see Table S4). Indeed, the carbonyl bond is significantly more contracted with {\CC{3}}/{\AVTZ}  (\InAA{1.208}) than with MP2/6-31G(d) (\InAA{1.221}). 
In particular, for the hallmark $n \rightarrow \pis$, our best estimate is {\IneV{$3.97$} (\emph{vide infra}), nicely matching a previous MR-AQCC value of \IneV{$3.98$}, \cite{Mul01} but significantly below the previous DMC/BLYP 
estimate of \IneV{$4.24$}. \cite{Sch04e} The latter discrepancy is probably due to the use of both different structures and pseudo-potentials within DMC calculations.
 
For methanimine and thioformaldehyde, the basis set effects are rather small for the states considered here (see Table S4) and the data reported in the present work are probably the most accurate vertical transition energies reported to date.
For the latter molecule, these vertical estimates are systematically larger than the known experimental 0-0 energies, \cite{Pao84} which is the expected trend.

\subsection{Larger compounds}

\begin{table}[htp]
\caption{\small Vertical (absorption) transition energies for various excited states of diazomethane (top) and ketene  (bottom). All values are in eV.}
\label{Table-5}
  \begin{footnotesize}
\begin{tabular}{ll|ccc|ccc|cc}
\hline 
		&		& \multicolumn{3}{c}{\AVDZ} & \multicolumn{3}{c}{\AVTZ} & \multicolumn{2}{c}{Litt.}\\
Molecule &State 	& {\CC{3}} & {\CCSDT}  &  {\exCI} & {\CC{3}} & {\CCSDT}  & {\exCI}& Exp. & Theo. \\
\hline
Acetaldehyde	&$^1A'' (n \rightarrow \pis)$			&4.34&4.32&4.34&	4.31&4.29&4.31	& 4.27$^a$ &4.29$^b$	\\
			&$^3A'' (n \rightarrow \pis)$			&3.96&3.95&3.98&	3.95&3.94&4.0$^c$  & 3.97$^a$ &3.97$^b$	\\
\hline
Cyclopropene	&$^1B_1 (\sigma \rightarrow \pis)$		&6.72&6.71&6.7$^c$&6.68&6.68&6.6$^c$& 6.45$^d$ & 6.89$^e$	\\
			&$^1B_2 (\pi \rightarrow \pis)$			&6.77&6.78&6.82&	6.73&6.75&6.7$^c$	& 7.00$^f$  & 7.11$^e$	\\
			&$^3B_2 (\pi \rightarrow \pis)$			&4.34&4.35&4.35&	4.34&	&4.38	& 4.16$^f$  & 4.28$^g$	\\
			&$^3B_1 (\sigma \rightarrow \pis)$		&6.43&6.43&6.43&	6.40&	&6.45	&		  & 6.40$^g$	\\
\hline%
Diazomethane	&$^1A_2 (\pi \rightarrow \pis)$ 			&3.10&3.10&3.09&	3.07&3.07&3.14	& 3.14$^h$ &3.21$^i$	\\
			&$^1B_1 (\pi \rightarrow 3s)$ 			&5.32&5.35&5.35&	5.45&5.48&5.54	&		   &5.33$^i$	\\
			&$^1A_1 (\pi \rightarrow \pis)$			&5.80&5.82&5.79&	5.84&5.86&5.90	& 5.9$^h$	   &5.85$^i$	\\
			&$^3A_2 (\pi \rightarrow \pis)$ 			&2.84&2.84&2.81&	2.83&2.82&2.8$^c$	&		   &2.92$^j$	\\
			&$^3A_1 (\pi \rightarrow \pis)$			&4.05&4.04&4.03&	4.03&4.02&4.05	&		   &3.97$^j$	\\
			&$^3B_1 (\pi \rightarrow 3s)$ 			&5.17&5.20&5.18&	5.31&5.34&5.35	&		   &			\\
			&$^3A_1 (\pi \rightarrow 3p)$			&6.83&6.83&6.81&	6.80&       &6.82	&		   &7.02$^j$	\\
			&$^1A'' [\mathrm{F}]	(\pi \rightarrow \pis)$	&0.68&0.67&0.65&	0.68&0.67&0.71	&	 	   &			\\
\hline
Formamide	&$^1A'' (n \rightarrow \pis)$			&5.71&5.68&5.70	&5.66&5.63&5.7$^c$&5.8$^k$		&5.63$^l$	\\
			&$^1A' (n \rightarrow 3s)$				&6.65&6.64&6.67	&6.74&6.74&		&6.35$^k$	&6.62$^l$	\\	
			&$^1A' (\pi \rightarrow \pis)$$^m$		&7.63&7.62&7.64	&7.62&	&7.63	&7.37$^k$	&7.22$^l$	\\
			&$^1A' (n \rightarrow 3p)$$^m$			&7.31&7.29&		&7.40&7.38&		&7.73$^k$	&7.66$^l$	\\
			&$^3A'' (n \rightarrow \pis)$			&5.42&5.39&5.42	&5.38&	&5.4$^c$	&5.2$^k$		&5.34$^l$	\\
			&$^3A' (\pi \rightarrow \pis)$			&5.83&5.81&5.82	&5.82&	&5.7$^c$	&$\sim$6$^k$	&5.74$^l$	\\
\hline%
Ketene		&$^1A_2 (\pi \rightarrow \pis)$ 			&3.89&3.88&3.84&	3.88&3.87&3.86	&3.7$^n$	&3.74$^o$	\\
			&$^1B_1 (n \rightarrow 3s)$ 			&5.83&5.86&5.88&	5.96&5.99&6.01	&5.86$^n$&5.82$^o$	\\
			&$^1A_2 (\pi \rightarrow 3p)$ 			&7.05&7.09&7.08&	7.16&7.20&7.18	&		&7.00$^o$	\\
			&$^3A_2 (n \rightarrow \pis)$ 			&3.79&3.78&3.79&	3.78&3.78&3.77	&3.8$^p$	&3.62$^q$\\
			&$^3A_1 (\pi \rightarrow \pis)$ 			&5.62&5.61&5.64&	5.61&5.60&5.61	&5$^p$	&5.42$^q$\\
			&$^3B_1 (n \rightarrow 3s)$ 			&5.63&5.66&5.68&	5.76&5.80&5.79	&5.8$^p$	&5.69$^q$\\
			&$^3A_2 (\pi \rightarrow 3p)$ 			&7.01&7.05&7.07&	7.12&7.17&7.12	&		&		\\
			&$^1A''[\mathrm{F}]	(\pi \rightarrow \pis)$	&1.00&0.99&0.96&	1.00&1.00&1.00	&		&	\\
\hline
Nitrosomethane&$^1A'' (n \rightarrow \pis)$			&2.00&1.98&1.99&	1.96&1.95&2.0$^c$	&1.83$^r$&1.76$^s$\\
			&$^1A' (n,n \rightarrow \pis,\pis)$		&5.75&5.26&4.81&	5.76&5.29	&4.72	&		&4.96$^s$\\
			&$^1A' (n \rightarrow 3s/3p)$ 			&6.20&6.19&6.29&	6.31&6.30&6.4$^c$	&		&6.54$^s$\\
			&$^3A'' (n \rightarrow \pis)$			&1.13&1.12&1.15&	1.14&1.13&1.16	&		&1.42$^t$\\
			&$^3A' (\pi \rightarrow \pis)$			&5.54&5.54&5.56&	5.51&	&5.60	&		&5.55$^t$\\
			&$^1A'' [\mathrm{F}]	(n \rightarrow \pis)$	&1.70&1.69&1.70&	1.69&1.66	&1.7$^c$	&		&	\\
\hline																			
Streptocyanine-C1&$^1B_2 (\pi \rightarrow \pis)$ 		&7.14&7.12&7.14&	7.13&7.11&7.1$^c$   &		&7.16$^u$\\
			&	$^3B_2 (\pi \rightarrow \pis)$ 		&5.48&5.47&5.47&	5.48&5.47&5.52	&		&		\\
\hline																			
 \end{tabular}
  \end{footnotesize} 
\begin{flushleft}
\begin{footnotesize}
$^a${Electron impact experiment from Ref.~\citenum{Wal87};}
$^b${NEVPT-PC from Ref.~\citenum{Ang05b};}
$^c${{\CI} convergence too slow to provide reliable estimates;}
$^d${Maximum in the gas UV from Ref.~\citenum{Rob69};}
$^e${CCSDT/TZVP from Ref.~\citenum{Kan14}; }
$^f${Electron impact experiment from Ref.~\citenum{Sau76};}
$^g${CC3/{\AVTZ} from Ref.~\citenum{Sil10c}; }
$^h${VUV maxima from Ref.~\citenum{McG71};}
$^i${{\CCSD}/6-311(3+,+)G(d) calculations from Ref.~\citenum{Fed07};}
$^j${MR-CC/DZP calculations from Ref.~\citenum{Rit89};}
$^k${EELS (singlet) and trapped electron (triplet) experiments from Ref.~\citenum{Gin97};}
$^l${$n$R-SI-CCSD(T) results from Ref.~\citenum{Li11}; }
$^m${Strong state mixing;}
$^n${Electron impact experiment from Ref.~\citenum{Fru76};}
$^o${CASPT2/6-311+G(d) results from Ref.~\citenum{Xia13};}
$^p${Electron impact experiment from Ref.~\citenum{Rob85b};}
$^q${STEOM-CCSD/Sad+//CCSD/Sad+ results from Ref.~\citenum{Noo03}.}
$^r${Maximum in the gas UV from Ref.~\citenum{Dix65};}
$^s${CASPT2/ANO results from Ref.~\citenum{Are06};}
$^t${CASSCF/cc-pVDZ results from Ref.~\citenum{Dol04};}
$^u${exCC3//MP2 result from Ref.~\citenum{Sen11}.}
\end{footnotesize}
\end{flushleft}
\end{table}

Let us now turn our attention to molecules that encompass three heavy (non-hydrogen) atoms. We have treated seven molecules of that family, and all were previously investigated at several levels of theory: acetaldehyde, 
\cite{Had93,Gwa95,Wib98,Ang05b,Rei09,Car10,Hoy16,Jac17b} cyclopropene, \cite{Sch08,She09b,Sil10b,Sil10c,Coe13,Kan14} diazomethane, \cite{Rit89,Hab95,Fed07,Rei09} formamide, \cite{Ser96,Bes99,Sch08,Sil10b,Sil10c,Kan14,Kan17} 
ketene, \cite{Rit89,Sza96b,Noo03,Xia13} nitrosomethane, \cite{Lac00,Dol04,Dol04b,Are06} and the shortest streptocyanine.\cite{Sen11,Bar13,Bou14,Zhe14,Leg15} The results are gathered in Table \ref{Table-5}. 
{Note that, for these molecules containing three heavy atoms, it is sometimes challenging to obtain reliable exFCI estimes, especially for the largest basis set.}

Experimentally, the lowest singlet and triplet $n \rightarrow \pis$ transitions of acetaldehyde are located \IneV{$0.3$--$0.4$} above their formaldehyde counterparts,\cite{Rob85b,Wal87} and this trend is accurately reproduced 
by theory, which also delivers estimates very close to the NEVPT2 values given in Ref.~\citenum{Ang05b}. 

For cyclopropene, the lowest singlet $\sigma \rightarrow \pis$ and $\pi \rightarrow \pis$ are close from one another, and both {\CCSDT} and {\exCI} predict the former to be slightly more stabilized, which is consistent with the 
large basis set {\CC{3}} results obtained previously by Thiel. \cite{Sil10c}

For the isoelectronic diazomethane and ketene molecules (see Table \ref{Table-5}), one notes, yet again, consistent results with, however, differences between the {\exCI}/{\AVTZ} and {\CCSDT}/{\AVTZ} results larger than \IneV{$0.05$} 
for the two lowest singlet states of diazomethane.  There is also a reasonable match between our data and previous theoretical results reported for these two molecules. \cite{Rit89,Noo03,Fed07,Xia13} The basis set effects are 
significant for the Rydberg transitions, especially for the $\pi \rightarrow 3s$ states of diazomethane (Table S5). 

In formamide, we found strong state mixing between the lowest singlet valence and Rydberg states of $A'$ symmetry. This is consistent with the  {\CCSDT}/TZVP analysis of Kannar and Szalay, \cite{Kan14} who reported, 
for example, a larger oscillator strength for the lowest Rydberg state than for the $\pi \rightarrow \pis$ transition.  This state-mixing problem pertains with {\AVTZ}, making unambiguous assignments impossible. 
Consequently, we have decided to classify the three lowest $^1A'$ transitions according to their dominant orbital character, which gives a picture consistent with the computed oscillator strengths (\emph{vide infra}) 
but yields state inversions compared to Thiel's and Szalay's assignments. \cite{Sil10b,Kan14} This strong state mixing also prevented the convergence of several state energies with the {\exCI}/{\AVTZ} approach. Despite these 
uncertainties, we obtained transition energies for the Rydberg states that are much closer from experiment \cite{Gin97} as well as from previous multireference {\CC{}} estimates, \cite{Li11} than the TZVP ones. \cite{Kan14}

Nitrosomethane is an interesting test molecule for three reasons: i) it presents very low-lying $n \rightarrow \pis$ states of $A''$ symmetry, close to ca.~\IneV{$2.0$} (singlet) and \IneV{$1.2$} (triplet), amongst the smallest absorption 
energies found in a compact molecule; \cite{Tar54} ii) it changes from an eclipsed to a staggered conformation of the methyl group when going from the ground to the lowest singlet state; \cite{Ern78,Gor79b,Dol04} iii) the lowest-lying 
singlet $A'$ state corresponds to an almost pure double excitation of $(n,n) \rightarrow (\pis,\pis)$ nature. \cite{Are06} Indeed, {\CC{3}} returns a $2.5$\%\ single excitation character only for this second transition, to be compared to 
more than $80$\%\ (and generally more than $90$\%) in all other states treated in this  work (\emph{vide infra}). For example, the notoriously difficult $A_g$ dark state of butadiene has a $72.8$\%\ single character. \cite{Sch08}
For the  $A''$ state of nitrosomethane, {\CC{3}}, {\CCSDT} and {\exCI} yield similar results, and the corresponding transition energies are slightly larger than previous {\CASPT} estimates. \cite{Are06}  In contrast, the {\CC{}} approaches 
are expectedly far from the spot for the $(n,n) \rightarrow (\pis,\pis)$ transition: they yield values significantly blue shifted and large discrepancies between the {\CC{3}} and {\CCSDT} values are found. For this particular state, it is not surprising 
that the {\exCI} result is indeed closer to the {\CASPT} value, \cite{Are06} as modeling double excitations with single-reference {\CC{}} models is not a natural choice.

Finally for the shortest model cyanine, a molecule known to be difficult to treat with {\TDDFT}, \cite{Leg15} all the theoretical results given in Table \ref{Table-5} closely match each other for both the singlet and triplet manifolds. 
For the former, the reported {\CASPT} (with IPEA) value of \IneV{$7.14$} also fits these estimates. \cite{Sen11}

\subsection{Theoretical best estimates}

We now turn to the definition of theoretical best estimates. We decided to provide two sets for these estimates, one obtained in the frozen-core approximation with the {\AVTZ} atomic basis set, and
one including further corrections for basis set and ``all electron'' (full) effects. This  choice allows further benchmarks to either consider a reasonably compact basis set, therefore allowing to test many levels 
of theory, or to rely on values closer to the basis set limit. {For the former set, we systematically selected} {\exCI}/{\AVTZ} {values except when explicitly stated.}
{For the latter set, both the ``all electron'' correlation and the basis set corrections (see} {\SI} {for complete data) were systematically obtained} at the {\CC{3}} level of theory and used {\DAVPZ} for the nine smallest 
molecules and slightly more compact basis sets for the larger compounds. At least for Rydberg states, the use of {\DAVQZ} apparently delivers results closer to basis set convergence than {\AVPZ}, and the
former basis set was used when technically possible.  {The interested readers may find in}  {\SI} {the values obtained with and without applying the frozen-core approximation
for several basis sets. Clearly, the largest amount of the total correction originates from basis set effects. In other words, ``full'' and frozen-core transition energies are typically within 0.01--0.02 eV of each
other for a given basis set.} The results are listed in Table \ref{Table-6} and provide a total of 110 transition energies. This set of states is rather diverse with 61 singlet and 45 triplet states, 
60 valence and 45 Rydberg states, 21 $n \rightarrow \pis$ and 38  $\pi \rightarrow \pis$ states, with an energetic span from $0.70$ to \IneV{$13.27$}. Amongst these 110 excitation energies, only 13 are characterized 
by a single-excitation character smaller than 90\%\ according to {\CC{3}}. As expected,\cite{Sch08} the dominant single-excitation character is particularly pronounced for triplet excited states. Therefore, this set 
is adequate for evaluating single-reference methods, though a few challenging cases are incorporated.  Consequently, we think that the TBE listed in Table \ref{Table-6} contribute to fulfill the need of more 
accurate reference excited state energies, as pointed out by Thiel one decade ago. \cite{Sch08} However, the focus on small compounds and the lack of charge-transfer states constitute significant biases 
in that set of transition energies.
 
 \bigskip

\renewcommand*{\arraystretch}{.55}
\LTcapwidth=\textwidth

\begin{footnotesize}
\begin{longtable}{llcccccc}
\caption{\small TBE (in eV) for various states and wave function approaches. For each state, we provide the oscillator strength and percentage of single excitations obtained at the \CC{3}(FC)/{\AVTZ} level. 
Unless otherwise stated, the TBE(FC)/{\AVTZ} have been obtained directly from {\exCI}. For the basis-set-corrected TBE, we provide the method used to determine the starting value and the basis set used at 
the \CC{3}(full) level to correct it. \CC{3}(full)/{\AVTZ} geometries and abbreviated forms of Dunning's basis set are systematically used.} \label{Table-6}\\
\hline 
			&		&	&		& TBE(FC)&  \multicolumn{3}{c}{Corrected TBE} \\
			& State	 & $f$ & \%$T_1$ & 	AVTZ	& Method & Corr.	& Value \\
\hline
\endfirsthead
\hline 
			&		&	&		& 	TBE(FC)&  \multicolumn{3}{c}{Corrected TBE} \\
			& State	 & $f$ & \%$T_1$ & AVTZ	&  Method & Corr.	& Value \\
\hline
\endhead
\hline \multicolumn{7}{r}{{Continued on next page}} \\
\endfoot
\hline
\endlastfoot
Acetaldehyde	&$^1A''(\mathrm{V};n \rightarrow \pis)$					& 0.000	&91.3& 4.31		& {\exCI}/AVTZ & AVQZ		&4.31	\\
			&$^3A''(\mathrm{V};n \rightarrow \pis)$					&		&97.9& 3.97$^a$ 	& {\exCI}/AVDZ & AVQZ		&3.98	\\
Acetylene		&$^1\Sigma_u^- (\mathrm{V};\pi \rightarrow \pis)$ 			&		&96.5& 7.10		& {\exCI}/AVTZ & dAV5Z	 	&7.10 	\\
			&$^1\Delta_u	(\mathrm{V};\pi \rightarrow \pis)$ 			&		&93.3& 7.44		&			&			&7.44 	\\	
			&$^3\Sigma_u^+ (\mathrm{V};\pi \rightarrow \pis)$ 			&		&99.2& 5.53		&			&			&5.56 	\\	
			&$^3\Delta_u	(\mathrm{V};\pi \rightarrow \pis)$ 			&		&99.0& 6.40		&			&			&6.40 	\\	
			&$^3\Sigma_u^- (\mathrm{V};\pi \rightarrow \pis)$ 			&		&98.8& 7.08		&			&			&7.09 	\\	
			&$^1A_u [\mathrm{F}]	(\mathrm{V};\pi \rightarrow \pis)$	&		&95.6& 3.64		&			&			&3.63 	\\	
			&$^1A_2 [\mathrm{F}]	(\mathrm{V};\pi \rightarrow \pis)$	&		&95.5& 3.85		&			&			&3.85	\\	
Ammonia		&$^1A_2 (\Ryd;n \rightarrow 3s)$ 						& 0.086	&93.5& 6.59		&{\exCI}/AVQZ & dAV5Z		&6.66 	\\
			&$^1E (\Ryd;n \rightarrow 3p)$ 							& 0.006	&93.7& 8.16		&			&			&8.21	\\	
			&$^1A_1 (\Ryd;n \rightarrow 3p)$ 						& 0.003	&94.0& 9.33		&			&			&8.65 	\\
			&$^1A_2 (\Ryd;n \rightarrow 4s)$ 						& 0.008	&93.6& 9.96		&{\exCI}/AVTZ  & dAV5Z		&9.19 	\\
			&$^3A_2 (\Ryd;n \rightarrow 3s)$ 						&		&98.2& 6.31		&{\exCI}/AVQZ & dAV5Z		&6.37	\\
Carbon monoxyde	&$^1\Pi (\mathrm{V};n \rightarrow \pis)$ 				& 0.084	&93.1 & 8.49		& {\exCI}/AVQZ& dAV5Z		&8.48	\\
			&$^1\Sigma^- (\mathrm{V};\pi \rightarrow \pis)$				&		&93.3 & 9.92		&			&			&9.98 	\\
			&$^1\Delta (\mathrm{V};\pi \rightarrow \pis)$ 				&		&91.8 &10.06		&			&			&10.10 	\\
			&$^1\Sigma^+ (\Ryd)$ 								& 0.003	&91.5 &10.95		&			&			&10.80 	\\
			&$^1\Sigma^+ (\Ryd)$ 								& 0.200	&92.9 &11.52		&			&			&11.42 	\\
			&$^1\Pi (\Ryd)$										& 0.053	&92.4 &11.72		&			&			&11.55 	\\
			&$^3\Pi (\mathrm{V};n \rightarrow \pis)$ 					&		&98.7 & 6.28		&			&			&6.28 	\\
			&$^3\Sigma^+ (\mathrm{V};\pi \rightarrow \pis)$			&		&98.7 & 8.45		&			&			&8.49 	\\
			&$^3\Delta (\mathrm{V};\pi \rightarrow \pis)$ 				&		&98.4 & 9.27		&			&			&9.28 	\\
			&$^3\Sigma^- (\mathrm{V};\pi \rightarrow \pis)$				&		&97.5 & 9.80		&			&			&9.77	\\
			&$^3\Sigma^+ (\Ryd)$ 								&		&98.0 & 10.47		&  {\exCI}/AVTZ &dAV5Z		&10.37 	\\
Cyclopropene	&$^1B_1 (\mathrm{V};\sigma \rightarrow \pis)$				& 0.001	&92.8 &6.68$^b$	& {\CCSDT}/AVTZ&AVQZ		& 6.68 	\\
			&$^1B_2 (\mathrm{V};\pi \rightarrow \pis)$				& 0.071	&95.1 &6.79$^a$	& {\exCI}/AVDZ	    &AVQZ		& 6.78 	\\
			&$^3B_2 (\mathrm{V};\pi \rightarrow \pis)$				&		&98.0 &4.38		& {\exCI}/AVTZ     &AVQZ 	& 4.38	 \\
			&$^3B_1 (\mathrm{V};\sigma \rightarrow \pis)$				& 		&98.9 &6.45		& 			    &			& 6.45 	\\
Diazomethane	&$^1A_2 (\mathrm{V};\pi \rightarrow \pis)$ 				&		&90.1 &3.14		&{\exCI}/AVTZ & dAVQZ		&3.13 	\\
			&$^1B_1 (\Ryd;\pi \rightarrow 3s)$ 						& 0.002	&93.8 &5.54		&			&			&5.59 	\\
			&$^1A_1 (\mathrm{V};\pi \rightarrow \pis)$				& 0.210	&91.4 &5.90		&			&			& 5.89	 \\
			&$^3A_2 (\mathrm{V};\pi \rightarrow \pis)$ 				&		&97.7 &2.79$^a$	& {\exCI}/AVDZ & dAVQZ		&2.80	\\
			&$^3A_1 (\mathrm{V};\pi \rightarrow \pis)$				&		&98.6 &4.05		& {\exCI}/AVTZ & dAVQZ		&4.05	\\
			&$^3B_1 (\Ryd;\pi \rightarrow 3s)$ 						&		&98.0 &5.35		&			&			&5.40 	\\
			&$^3A_1 (\Ryd;\pi \rightarrow 3p)$						&		&98.5 &6.82		&			&			&6.72 	\\
			&$^1A'' [\mathrm{F}]	(\mathrm{V};\pi \rightarrow \pis)$		&		&87.4 &0.71		& 			 & 			& 0.70 	\\
Dinitrogen		&$^1\Pi_g (\mathrm{V};n \rightarrow \pis)$ 				&		&92.6 &9.34		& {\exCI}/AVQZ&dAV5Z		&9.33	 \\
			&$^1\Sigma_u^- (\mathrm{V};\pi \rightarrow \pis)$			&		&97.2 &9.88		&			&			&9.91	 \\
			&$^1\Delta_u (\mathrm{V};\pi \rightarrow \pis)$ 				& 0.000	&95.9 &10.29		&			&			&10.31	 \\
			&$^1\Sigma_g^+ (\Ryd)$ 								&		&92.2 &12.98		&			&			&12.30	 \\
			&$^1\Pi_u (\Ryd)$ 									& 0.229	&82.9 &13.03		& {\exCI}/AVTZ&dAV5Z		&12.73	 \\
			&$^1\Sigma_u^+ (\Ryd)$ 								& 0.296	&92.8 &13.09		&			&			&12.95	 \\
			&$^1\Pi_u (\Ryd)$ 									& 0.000	&87.4 &13.46		&			&			&13.27	 \\
			&$^3\Sigma_u^+ (\mathrm{V};\pi \rightarrow \pis)$			&		&99.3 &7.70		& {\exCI}/AVQZ&dAV5Z		&7.74	 \\
			&$^3\Pi_g (\mathrm{V};n \rightarrow \pis)$ 				&		&98.4 &8.01		&			&			&8.03	 \\
			&$^3\Delta_u (\mathrm{V};\pi \rightarrow \pis)$ 				&		&99.3 &8.87		&			&			&8.88	 \\
			&$^3\Sigma_u^- (\mathrm{V};\pi \rightarrow \pis)$			&		&98.8 &9.66		&			&			&9.65	 \\
Ethylene		&$^1B_{3u} (\Ryd;\pi \rightarrow 3s)$ 					& 0.078	&95.1 &7.39		&{\exCI}/AVTZ & dAV5Z		&7.43	 \\
			&$^1B_{1u} (\mathrm{V};\pi \rightarrow \pis)$ 				& 0.346	&95.8 &7.93		&			& 			&7.92	\\
			&$^1B_{1g} (\Ryd;\pi \rightarrow 3p)$ 					&		&95.3 &8.08		& 			& 			&8.10	\\
			&$^3B_{1u} (\mathrm{V};\pi \rightarrow \pis)$ 				&		&99.1 &4.54		& 			&			&4.54	\\
			&$^3B_{3u} (\Ryd;\pi \rightarrow 3s)$ 					&		&98.5 &7.23$^a$	&{\exCI}/AVDZ& dAV5Z		&7.28	\\
			&$^3B_{1g} (\Ryd;\pi \rightarrow 3p)$ 					&		&98.4 &7.98$^a$	&			&			&8.00	\\
Formaldehyde	&$^1A_2 (\mathrm{V}; n \rightarrow \pis)$ 				&		&91.5 &3.98		&{\exCI}/AVTZ & dAV5Z		&3.97	\\
			&$^1B_2 (\Ryd;n \rightarrow 3s)$ 						& 0.021	&91.7 &7.23		&			 &			&7.30	\\
			&$^1B_2 (\Ryd;n \rightarrow 3p)$ 						& 0.037	&92.4 &8.13		&			&			&8.14	\\
			&$^1A_1 (\Ryd;n \rightarrow 3p)$ 						& 0.052	&91.9 &8.23		&			&			&8.27	\\
			&$^1A_2 (\Ryd;n \rightarrow 3p)$ 						&		&91.7 &8.67		&			&			&8.50	\\
			&$^1B_1 (\mathrm{V};\sigma \rightarrow \pis)$				& 0.001	&90.8 &9.22		&			&			&9.21	\\
			&$^1A_1 (\mathrm{V};\pi \rightarrow \pis)$				& 0.135	&90.4 &9.43		&			&			&9.26	\\
			&$^3A_2 (\mathrm{V};n \rightarrow \pis)$ 					&		&98.1 &3.58		&			&			&3.58	\\
			&$^3A_1 (\mathrm{V};\pi \rightarrow \pis)$				&		&99.0 &6.06		&			&			&6.07	\\
			&$^3B_2 (\Ryd;n \rightarrow 3s)$ 						&		&97.1 &7.06		&			&			&7.14	\\
			&$^3B_2 (\Ryd;n \rightarrow 3p)$ 						&		&97.4 &7.94		&			&			&7.96	\\
			&$^3A_1 (\Ryd;n \rightarrow 3p)$ 						&		&97.2 &8.10		&			&			&8.15	\\
			&$^3B_1 (\Ryd;n \rightarrow 3d)$ 						&		&97.9 &8.42		&			&			&8.42	\\
			&$^1A'' [\mathrm{F}] (\mathrm{V};n \rightarrow \pis)$			&		&87.8 &2.80		&			&			&2.80	\\
Formamide	&$^1A'' \mathrm{V};(n \rightarrow \pis)$					&0.000	&90.8 &5.65$^a$	&{\exCI}/AVDZ& AVQZ		&5.63	\\
			&$^1A' (\Ryd;n \rightarrow 3s)$							&0.001	&88.6 &6.77$^a$	&			&			&6.81	\\
			&$^1A' (\mathrm{V};\pi \rightarrow \pis)$					&0.251	&89.3 &7.63		&{\exCI}/AVTZ & AVQZ		&7.64	\\
			&$^1A' (\Ryd;n \rightarrow 3p)$							&0.111	&89.6 &7.38$^b$	&{\CCSDT}/AVTZ& AVQZ		&7.41	\\
			&$^3A'' (\mathrm{V};n \rightarrow \pis)$					&		&97.7 &5.38$^c$	&{\exCI}/AVDZ& AVQZ		&5.37	\\
			&$^3A' (\mathrm{V};\pi \rightarrow \pis)$					&		&98.2 &5.81$^c$	&			&			&5.81	\\
Hydrogen chloride	 & $^1\Pi (\mathrm{CT})$							&0.056	&94.3 &7.84		& {\exCI}/AVQZ &dAV5Z		&7.86	\\	
Hydrogen sulfide &$^1A_2 (\Ryd;n \rightarrow 4p)$ 						&		&94.6 &6.18		& {\exCI}/AVQZ &dAV5Z		&6.10	\\
			&$^1B_1 (\Ryd;n \rightarrow 4s)$ 						& 0.063	&94.3 &6.24		&  			&			&6.29	\\
			&$^3A_2 (\Ryd;n \rightarrow 4p)$ 						&		&98.7 &5.81		&  			&			&5.74	\\
			&$^3B_1 (\Ryd;n \rightarrow 4s)$ 						&		&98.4 &5.88		& 			&			&5.94	 \\
Ketene		&$^1A_2 (\mathrm{V};\pi \rightarrow \pis)$ 				&		&91.0 &3.86		&{\exCI}/AVTZ & dAVQZ		&3.86	\\
			&$^1B_1 (\Ryd;n \rightarrow 3s)$ 						& 0.035	&93.9 &6.01		&			&			&6.06	\\
			&$^1A_2 (\Ryd;\pi \rightarrow 3p)$ 						&		&94.4 &7.18		&			&			&7.19	\\
			&$^3A_2 (\mathrm{V};n \rightarrow \pis)$ 					&		&91.0 &3.77		&			&			&3.77	\\
			&$^3A_1 (\mathrm{V};\pi \rightarrow \pis)$ 				&		&98.6 &5.61		&			&			&5.60	\\
			&$^3B_1 (\Ryd;n \rightarrow 3s)$ 						&		&98.1 &5.79		&			&			&5.85	\\
			&$^3A_2 (\Ryd;\pi \rightarrow 3p)$ 						&		&94.4 &7.12		&			&			&7.14	\\
			&$^1A'' [\mathrm{F}] (\mathrm{V};\pi \rightarrow \pis)$		&		&87.9 &1.00		&			 & 			&1.00	\\
Methanimine	&$^1A''(\mathrm{V}; n \rightarrow \pis)$ 					&0.003	&90.7 &5.23		&{\exCI}/AVTZ & dAVQZ		&5.21	\\
			&$^3A'' (\mathrm{V}; n \rightarrow \pis)$					&		&98.1 &4.65		&			&			&4.64	\\	
Nitrosomethane&$^1A'' (\mathrm{V};n \rightarrow \pis)$					& 0.000	&93.0 &1.96$^a$	& {\exCI}/AVDZ & AVQZ		&1.95	 \\
			&$^1A' (\mathrm{V};n,n \rightarrow \pis,\pis)$				&0.000	&2.5	  &4.72		& {\exCI}/AVTZ & AVQZ		& 4.69	\\
			&$^1A' (\Ryd;n \rightarrow 3s/3p)$ 						&0.006	&90.8 &6.40$^a$	&{\exCI}/AVDZ	& AVQZ		&6.42 	\\
			&$^3A'' (\mathrm{V};n \rightarrow \pis)$					&		&98.4 &1.16		&			&			&1.16 	\\
			&$^3A' (\mathrm{V};\pi \rightarrow \pis)$					&		&98.9 &5.60		&			&			&5.61	Ê\\
			&$^1A'' [\mathrm{F}]	(\mathrm{V};n \rightarrow \pis)$			&		&92.7&1.67$^a$	&{\exCI}/AVDZ	& AVQZ		&1.66	\\
Streptocyanine-C1&$^1B_2 (\mathrm{V};\pi \rightarrow \pis)$ 				& 0.347	&88.7&7.13$^a$	& {\exCI}/AVDZ & AVQZ		&7.12	\\
			&	$^3B_2 (\mathrm{V};\pi \rightarrow \pis)$ 				& 		&98.3 &5.52		& {\exCI}/AVTZ & AVQZ		&5.52	 \\
Thioformaldehyde&$^1A_2 (\mathrm{V};n \rightarrow \pis)$ 				&		&89.3 &2.22		&  {\exCI}/AVTZ & dAVQZ		&2.20	\\
			&$^1B_2 (\Ryd;n \rightarrow 4s)$ 						& 0.012	&92.3 &5.96		&			&			&5.99	\\
			&$^1A_1 (\mathrm{V};\pi \rightarrow \pis)$				& 0.178	&90.8 &6.38$^d$	&{\CCSDTQ}/AVDZ& dAVQZ	&6.34	\\
			&$^3A_2 (\mathrm{V};n \rightarrow \pis)$ 					&		&97.7 &1.94		&{\exCI}/AVTZ & dAVQZ		&1.94	\\
			&$^3A_1 (\mathrm{V};\pi \rightarrow \pis)$				&		&98.9 & 3.43		&			&			&3.44	\\
			&$^3B_2 (\Ryd;n \rightarrow 4s)$ 						&		&97.6 &5.72$^a$	&{\exCI}/AVDZ& dAVQZ		&5.76	\\
			&$^1A_2 [\mathrm{F}] (\mathrm{V};n \rightarrow \pis)$		&		&87.2 &1.95		&{\exCI}/AVTZ & dAVQZ		&1.94	\\
Water		& $^1B_1 (\Ryd; n \rightarrow 3s)$ 						& 0.054	&93.4 &7.62		& {\exCI}/AVQZ&dAV5Z 		&7.70	 \\
			& $^1A_2 (\Ryd; n \rightarrow 3p)$ 						& 		&93.6 &9.41		&			&			&9.47 	 \\	
			& $^1A_1 (\Ryd; n \rightarrow 3s)$ 						& 0.100	&93.6 &9.99		&			&			&9.97	 \\	
			& $^3B_1 (\Ryd; n \rightarrow 3s)$ 						& 		&98.1 &7.25		&			&			&7.33	 \\	
			& $^3A_2 (\Ryd; n \rightarrow 3p)$ 						& 		&98.0 &9.24 		&			&			&9.30	 \\	
			& $^3A_1 (\Ryd; n \rightarrow 3s)$ 						& 		&98.2 &9.54		&			&			&9.59	 \\	
 \end{longtable}
  \end{footnotesize}
  \vspace{-1.0 cm}
\begin{flushleft}\begin{footnotesize}\begin{singlespace}
$^a${{exCI}/{\AVDZ} data corrected with the difference between {\CCSDT}/{\AVTZ} and {\CCSDT}/{\AVDZ} values;}
$^b${{\CCSDT}/{\AVTZ} value;}
$^c${{exCI}/{\AVDZ} data corrected with the difference between {\CC{3}}/{\AVTZ} and {\CC{3}}/{\AVDZ} values;}
$^d${{\CCSDTQ}/{\AVDZ} data corrected with the difference between {\CCSDT}/{\AVTZ} and {\CCSDT}/{\AVDZ} values.}
\end{singlespace}\end{footnotesize}\end{flushleft}

\subsection{Benchmarks}

We have used the TBE(FC)/{\AVTZ} benchmark values to assess the performances of twelve wavefunction approaches, namely,  {\ADC{2}}, {\ADC{3}}, CIS(D),  CIS(D$_\infty$), {\CC{2}}, STEOM-CCSD,  {\CCSD},  
CCSDR(3), CCSDT-3,  {\CC{3}}, {\CCSDT} and {\CCSDTQ}.  The complete list of results can be found in Table S6 in the {\SI}. As expected, only the approaches including iterative
triples, that is,  {\ADC{3}}, CCSDT-3,  {\CC{3}} and {\CCSDT} are able to predict the presence of the doubly excited $(n,n) \rightarrow (\pis,\pis)$ transition in nitrosomethane (see Tables \ref{Table-5} and S6), 
but they all yield large quantitative errors. Indeed, the TBE value of \IneV{4.72} is strongly underestimated by {\ADC{3}} (\IneV{3.00}) and significantly overshot by the three {\CC{}} models with 
estimates of \IneV{$6.02$}, \IneV{$5.76$} and \IneV{$5.29$} with CCSDT-3, {\CC{3}}, and {\CCSDT}, respectively. This \IneV{$0.26$} difference between the CCSDT-3 and {\CC{3}} values is also 
the largest discrepancy between these two models in the tested set.  Obviously, from a general perspective, one should not use the standard single-reference wavefunction methods to describe double excitations. 
Therefore, the $(n,n) \rightarrow (\pis,\pis)$ transition of nitrosomethane was removed from our statistical analysis. Likewise, for the three lowest $^1A'$ excited states 
of formamide, strong state mixing --- involving two or three states --- are found at all levels of theory, making unambiguous assignments impossible. Consequently, they are also excluded from our
 statistics. 

In Table \ref{Table-7}, we report, for the entire set of compounds, the mean signed error (MSE), mean absolute error (MAE) root mean square deviation (RMS), as well as the positive [\MaxP] and negative [\MaxN] maximum deviations.
A graphical representation of the errors obtained with all methods can be found in Figure \ref{Fig-1}. Note that only singlet states could be computed with the programs used for CCSDR(3) and CCSDT-3. 
As shown in Fig.~\ref{Fig-1}, {\CCSDTQ} is on the spot with tiny MSE and MAE, which is consistent with the analysis carried out for individual molecules.  With this method, the negative and positive 
maximum deviations are as small as \IneV{$-0.05$} (singlet $n \rightarrow 4s$ Rydberg transition of thioformaldehyde) and \IneV{$+0.06$}  ($^1\Sigma_u^+$  Rydberg transition of dinitrogen), respectively.
The three other {\CC{}} models with iterative triples ({\CCSDT}-3, {\CC{3}}, and {\CCSDT}) also deliver extremely accurate transition energies with MAE of \IneV{$0.03$} only. In agreement with the analysis 
of Watson and co-workers, we do not find any significant (statistical) differences between {\CCSDT}-3 and {\CC{3}}, \cite{Wat13} and although the former theory is formally closer to {\CCSDT}, it does not
seem more advantageous nor disadvantageous than {\CC{3}} in practice. The very good performance of {\CC{3}} is also consistent with the analysis of Thiel and coworkers, who reported a strong agreement 
with {\CASPT}, \cite{Sil10c}  as well as with the conclusion of Szalay's group who found it very close to {\CCSDT}. \cite{Kan17} Nevertheless, {\CCSDT} is not, on average, significantly more accurate than {\CC{3}} 
nor CCSDT-3. In other words, {\CCSDT} is probably not a sufficiently accurate benchmark to estimate the accuracy of {\CCSDT}-3 nor {\CC{3}}. The perturbative inclusion of triples via CCSDR(3) stands as a good 
compromise between computational cost and accuracy with a MAE of \IneV{$0.04$}, a conclusion also drawn in the benchmark study performed by Sauer and coworkers.\cite{Sau09}  These very small average 
deviations are related to the fact that the majority of our set is constituted of large single-excitation  character transitions (see \%$T_1$ in Table \ref{Table-6}). Reasonably, we predict that they would slightly deteriorate 
for larger compounds. 

\renewcommand*{\arraystretch}{1.0}
\begin{table}[htp]
\caption{Mean signed error (MSE), mean absolute error (MAE), root-mean square deviation (RMS), positive [\MaxP] and negative [\MaxN] maximal deviations with respect to TBE(FC)/{\AVTZ} for the transition energies listed in Table S6. 
All values are in eV and have been obtained with the {\AVTZ} basis set.}
\label{Table-7}
\begin{tabular}{lcccccc}
\hline 
Method & Nb. States 	& MSE & MAE & RMS & \MaxP & \MaxN \\
\hline
CIS(D)		&106		&0.10	&0.25	&0.32	&-0.63	&1.06	\\
CIS(D$_\infty$)	&106		&-0.01	&0.21	&0.28	&-0.76	&0.57	\\
{\CC{2}}		&106		&0.03	&0.22	&0.28	&-0.71	&0.63	\\
STEOM-CCSD	&102		&0.01	&0.10	&0.14	&-0.56	&0.40	\\
 {\CCSD}		&106		&0.05	&0.08	&0.11	&-0.17	&0.40	\\
CCSDR(3)	&59		&0.01	&0.04	&0.05	&-0.07	&0.25	\\
CCSDT-3		&58		&0.01	&0.03	&0.05	&-0.07	&0.24	\\
{\CC{3}}		&106		&-0.01	&0.03	&0.04	&-0.09	&0.19	\\
 {\CCSDT}		&104		&-0.01	&0.03	&0.03	&-0.10	&0.11	\\
 {\CCSDTQ}	&73		&0.00	&0.01	&0.02	&-0.05	&0.06	\\
{\ADC{2}}		&106		&-0.01	&0.21	&0.28	&-0.76	&0.57	\\
{\ADC{3}}		&106		&-0.15	&0.23	&0.28	&-0.79	&0.39	\\
\hline																			
 \end{tabular}
 \end{table}
   
For the second-order {\CC{}} series, as expected, the errors increase when one uses more approximate models. Indeed, the MAE are $0.08$, $0.10$, and \IneV{$0.22$} with {\CCSD}, STEOM-CCSD and {\CC{2}}, respectively. 
The magnitude of the {\CC{2}} average deviation is consistent with previous estimates obtained for Thiel's set (\IneV{0.29} for singlets and \IneV{0.18} for triplets), \cite{Sch08} for fluorescence energies (\IneV{0.21} for 12 small 
compounds),\cite{Jac18a} as well as for larger compounds (\IneV{0.15} for 0-0 energies of conjugated dyes). \cite{Jac15b}  Likewise, the fact that {\CCSD} tends to overestimate the transition energies (positive MSE) was also 
reported previously in several works. \cite{Sch08,Car10,Wat13,Kan14,Jac17b,Kan17,Jac18a} It can be seen that Nooijen's STEOM approach, which was much less benchmarked previously, delivers an accuracy comparable to {\CCSD}, 
with a smaller MSE but a large dispersion.  More surprisingly, we found a MAE  smaller with {\CCSD} than with {\CC{2}}, which contrasts with the results reported for Thiel's set, \cite{Sau09} but is consistent with 
Kannar, Tajti and  Szalay conclusion. \cite{Kan17} We attribute this effect to the small size of the  compounds treated herein. Indeed, analyzing the TZVP values of Ref.~\citenum{Sch08}, it appears clearly that 
{\CC{2}} more regularly outperforms {\CCSD} for larger compounds.

As expected, the results for CIS(D$_\infty$) and {\ADC{2}}, two closely related theories, \cite{Hat05c,Dre15} are nearly equivalent, with only 4 (out of 106) cases for which a difference of \IneV{0.01} could be evidenced (Table S6).
In addition, Table \ref{Table-7} evidences that {\ADC{2}} provides an accuracy similar to  {\CC{2}} for a smaller computational cost, whereas CIS(D) is slightly less accurate. Both outcomes perfectly fit previous benchmarks. 
\cite{Hat05c,Win13,Har14,Jac15b,Jac18a} Conversely, we found that {\ADC{3}} results are rather poor with average deviations larger than the ones obtained with {\ADC{2}} and a clear tendency to provide red-shifted transition energies 
with a MSE of \IneV{$-0.15$}. This observation is in sharp contrast with a previous investigation which concluded that {\ADC{3}} and {\CC{3}} have very similar performances, \cite{Har14} though the {\ADC{3}} excitation energies 
were also found to be, on average, smaller by \IneV{$0.20$} compared to} their {\CC{3}} counterparts. At this stage, it is difficult to know if the large MAE of {\ADC{3}} reported in Table \ref{Table-7} originates solely from the small 
size of the compounds treated herein. However, the fact that the {\CCSD} MSE is relatively small compared to previous benchmarks hints that the choice of compact compounds has a non-negligible effect on the statistics. 

\begin{figure}[htp]
  \includegraphics[scale=0.98,viewport=2cm 10cm 19cm 27.5cm,clip]{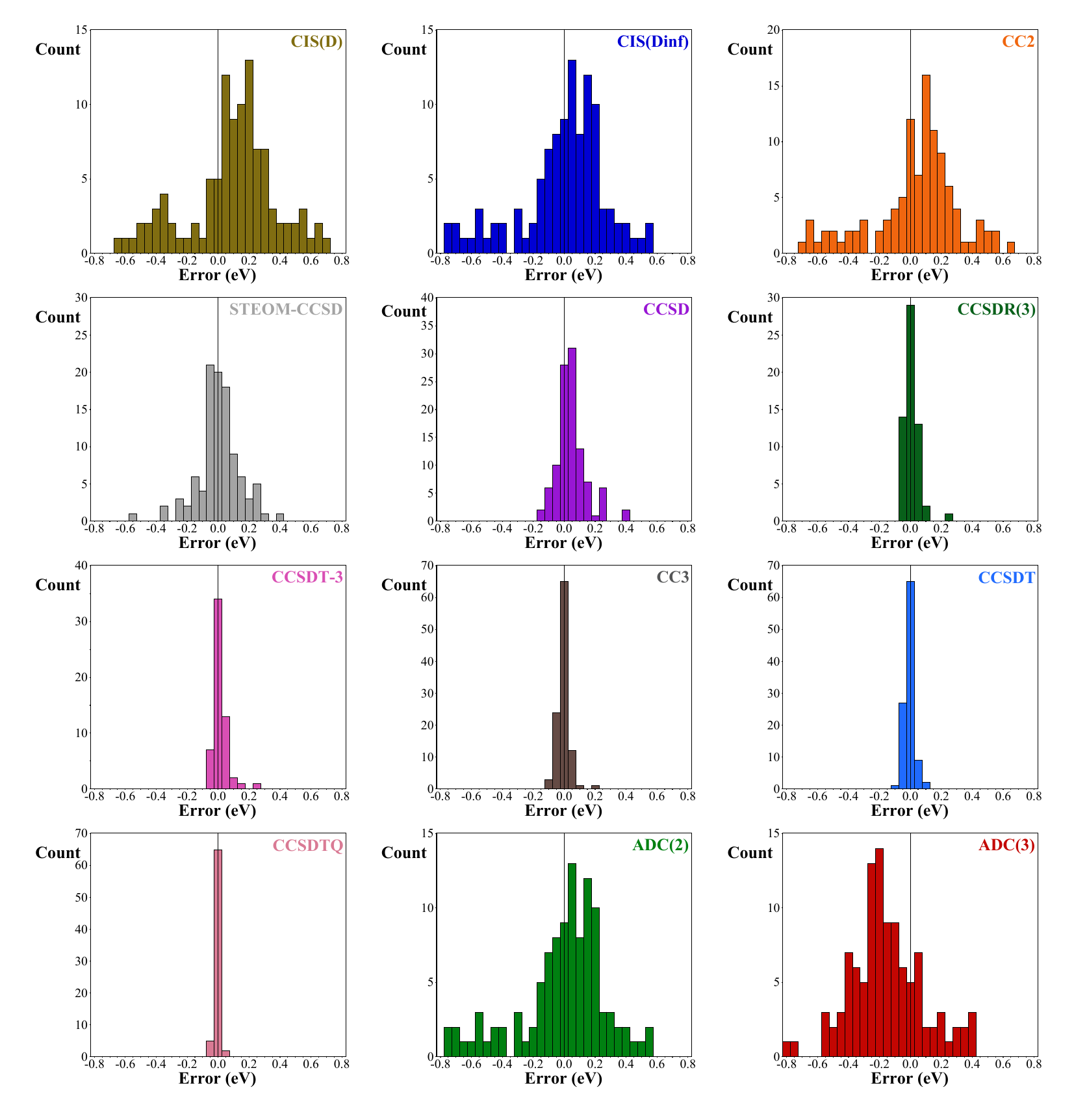}
  \caption{Histograms of the error patterns for several wavefunction methods compared to TBE(FC).  Note the variation of scaling of the vertical axes.}
  \label{Fig-1}
\end{figure}

\begin{figure}[htp]
  \includegraphics[scale=0.45,viewport=6cm 3cm 22cm 16cm,clip]{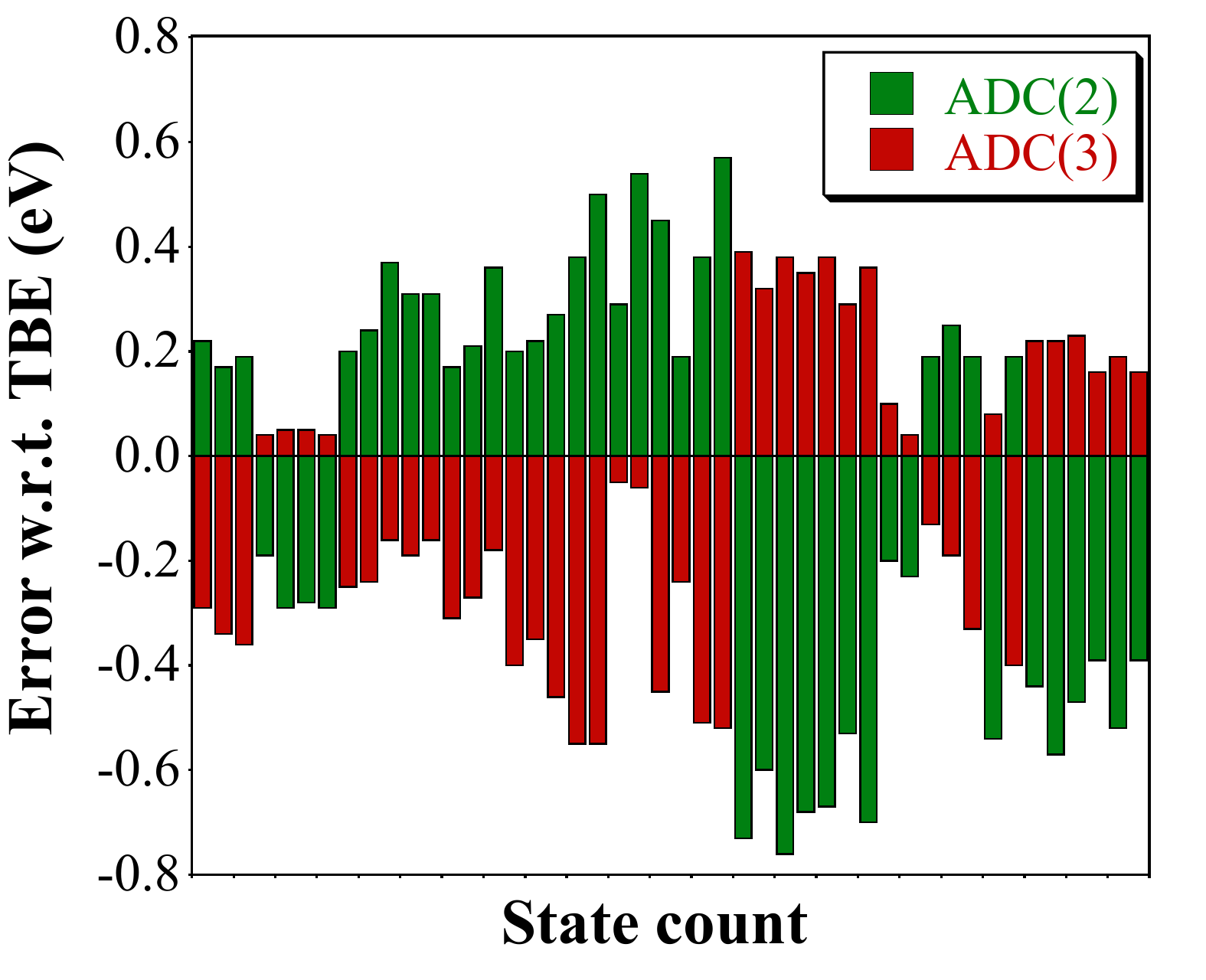}
  \caption{Comparison between the errors obtained with {\ADC{2}} and {\ADC{3}}  [compared to TBE(FC)] for the 46 states for which {\ADC{2}} yields an absolute deviation larger than \IneV{$0.15$}. 
  All values are in eV.}
  \label{Fig-2}
\end{figure}

Let us analyze the {\ADC{3}} errors more thoroughly. First, {\ADC{3}} deviations are quite large for all subsets (\emph{vide infra}). Second, we have found that, for the 46 transition energies for which {\ADC{2}} yields an absolute error exceeding 
\IneV{$0.15$} compared to our TBE, the signs of the  {\ADC{2}} and {\ADC{3}} errors systematically differ (see Figure \ref{Fig-2}), i.e., {\ADC{3}} goes in the right ``direction''  but has the tendency to over-correct {\ADC{2}}. 
This is clearly reminiscent of the well-known oscillating behavior of the M{\o}ller-Plesset perturbative series for ground state properties. Third, this overestimation of the corrections pertains for the states in which the {\ADC{2}} absolute error 
is smaller than \IneV{$0.15$}. Indeed, in those 60 cases, there are only 10 transitions for which the {\ADC{3}} values are more accurate than their second-order counterpart. As a consequence, taking the average between the {\ADC{2}} 
and {\ADC{3}}  transition energies yield rather accurate estimates with a MAE as small as \IneV{0.10} for the full set, half of the MAE obtained with the parent methods. 

We provide a more detailed analysis for several subsets of states in Table S7 in the SI. Globally, we found no significant difference between the singlet and triplet transitions, 
though all {\CC{} models (except STEOM-CCSD)} provide slightly smaller deviations for the latter transitions, in line with their larger single-excitation character. With the computationally lighter methods, CIS(D), CIS(D$_\infty$),
{\ADC{2}}, and {\CC{2}}, the MAEs are significantly smaller for the valence transitions (0.20, 0.15, 0.15, and \IneV{$0.18$}, respectively) than for the Rydberg transitions ($0.32$, $0.29$, $0.29$, and \IneV{$0.26$}, respectively).
We also found MSE of opposite signed for valence and Rydberg transitions with {\CC{2}}, which fits the results of Kannar and coworkers. \cite{Kan17} Surprisingly, {\ADC{3}} gives $0.28$ and \IneV{$0.17$} MAE for valence and Rydberg, respectively. 
All  {\CC{}} methods including triples  theories deliver similar deviations for both sets of states. All methods provide smaller (or equal) MAE for the $n \rightarrow \pis$ than for the $\pi \rightarrow \pis$ transitions, which was already found for 
Thiel's set.\cite{Sch08} The differences are particularly significant with CIS(D), {\CC{2}},  STEOM-CCSD and {\ADC{3}} with errors twice larger for $\pi \rightarrow \pis$ than $n \rightarrow \pis$ states. Finally, when considering the 
few states with \%$T_1$ smaller than 90\%, we logically found larger statistical errors  with, for example, MAE of, e.g., \IneV{$0.03$} for {\CCSDTQ}, \IneV{$0.04$} for {\CC{3}}, and \IneV{$0.06$} for CCSDT-3. 

\subsection{On the use of a compact basis set}

In several of the molecules considered here, we have found that adding corrections for basis set effects determined at the {\CC{3}} level to {\exCI}/{\AVDZ} results effectively provides accurate estimates of the {\exCI}
values directly determined with larger bases. Nevertheless, the dreadful scalings of both {\exCI} and {\CCSDTQ} make the size of the atomic basis the central bottleneck. For this reason, we have tested the use
of  one of most compact basis encompassing both diffuse and polarization functions, namely Pople's {\PopleDZ}. We have performed {\CC{3}}, {\CCSDT}, and {\CCSDTQ} calculations with this particular basis.  
The results are collected in the {\SI} (Table S8). First, we compare the {\PopleDZ} results to those obtained with the same theoretical method in conjunction with the {\AVTZ} basis set. As expected, large discrepancies 
are found with mean absolute deviation of $0.20$, $0.19$, and  \IneV{$0.25$}, for {\CC{3}}, {\CCSDT}, and {\CCSDTQ}, respectively. \cite{zzz-tou-2} Secondly, by adding the differences between the {\CC{3}}/{\AVTZ} and 
{\CC{3}}/{\PopleDZ} results to the {\CCSDT}/{\PopleDZ} and {\CCSDTQ}/{\PopleDZ} values, we obtained improved values. Such procedure yields very good estimates of the actual {\AVTZ} results, as the 
MAE are down to \IneV{0.01} with no error larger than \IneV{0.04} for both {\CCSDT} and {\CCSDTQ}. This is a particularly remarkable result for Rydberg states that are extremely basis set dependent.
For example, for the $^3A_2 (n \rightarrow 3p)$ transition in water, the {\CCSDTQ}/{\PopleDZ} value of \IneV{$10.34$} is more than \IneV{1} above its {\CCSDTQ}/{\AVTZ} counterpart (\IneV{$9.23$}, see Table \ref{Table-1}).
Applying the {\CC{3}} basis set correction makes the final error as small as \IneV{0.03}. This composite methodology opens the way to calculations on larger systems without significant loss of accuracy.

\section{Conclusions and outlook}

We have defined a set of more than 100 vertical transition energies, as close as possible to the {\FCI} limit. To this end, we have used both the coupled cluster route up to the highest computationally-possible 
order and the selected configuration interaction route up to the largest technically-affordable number of determinants, that is here about few millions. These calculations have been performed on 18 compounds encompassing one, two or 
three non-hydrogen atoms, using geometries optimized at the {\CC{3}} level and a series of diffuse Dunning's basis sets of increasing size. It was certainly gratifying to find extremely good agreements 
between the results obtained independently with these two distinct approaches with  typical differences as small as \IneV{$0.01$} between {\CCSDTQ} and {\exCI} transition energies. In fact, during the course 
of this joint work, the two groups involved in this study were able to detect misprints or incorrect assignments in each others calculations even when the differences were apparently negligible. 
For the two diatomic molecules considered in this work, \ce{N2} and \ce{CO}, the mean absolute deviation between our theoretical best estimates and the ``experimental'' vertical transition energies deduced 
from spectroscopic measurements using a numerical solution of the nuclear Schr\"odinger equation is as small as \IneV{$0.04$}, and it was possible to resolve previous inconsistencies between these 
``experimental'' values.  A significant share of the remaining error is likely related to the use of theoretically-determined geometries. Although, it is not possible to provide a definitive error bar for the 110 TBE 
listed in this work, our estimate, based on the differences between the two routes as well as the extrapolations used in the {\sCI}  procedure, is \IneV{$\pm 0.03$}. 

In another part of this work, we have used the TBE(FC)/{\AVTZ} values to benchmark a series of twelve popular wavefunction approaches. For the computationally most effective approaches, CIS(D), CIS(D$_\infty$),  {\ADC{2}}, 
and {\CC{2}}, we found average deviations of ca.~$0.21$--\IneV{$0.25$} with strong similarities between the {\ADC{2}} and {\CC{2}} results. Both conclusions are backed up by previous works. 
Likewise, we obtained the expected trend that {\CCSD} overestimates the transition energies, though with an amplitude that is quite small here, likely due to the small size of the compounds investigated. 
More interestingly, we could demonstrate that STEOM-CCSD is, on average, as accurate as {\CCSD}, and we were also able to benchmark the methods including contributions from triples using reliable 
theoretical references. Interestingly, we found no significant differences between CCSDT-3, {\CC{3}}, and {\CCSDT}, that all yield a MAE of \IneV{0.03}. In other words, we could not demonstrate that {\CCSDT} 
is statistically more accurate than its approximated (and computationally more effective) forms, nor highlight significant differences between CCSDT-3 and {\CC{3}}. We have observed that the use of perturbative triples, 
as in CCSDR(3), allows to correct most of the {\CCSD} error. This evidences that CCSDR(3) is a computationally appealing method as it gives average deviations only slightly larger than with iterative triples. In contrast, 
for the present set of molecules, {\ADC{3}} was found significantly less accurate than {\CC{3}}, and it was showed that {\ADC{3}} over-corrects {\ADC{2}}. Whether this surprising result is related to the size of the compounds 
or is a more general trend remains to be confirmed. 

As stated several times throughout this work, the size of the considered molecules is certainly one of the main limitations of the present effort, as it introduces a significant bias, e.g., charge-transfer over several {\AA} are totally 
absent of the set.  Obviously, the respective $\order*{N^{10}}$ and $\order*{e^N}$ formal scalings of {\CCSDTQ} and {\FCI} do not offer an easy pathway to circumvent this limit. Nevertheless, it appears that performing 
{\exCI} calculations with a relatively compact basis, e.g., {\AVDZ} or even {\PopleDZ}, and correcting the basis set effects with a more affordable approach, e.g., {\CC{3}}, might be a valuable and efficient approach to reach 
accurate vertical excitations energies for larger molecules, at least for the electronic transitions presenting a dominant single excitation character. Indeed, we have shown here that such basis set extrapolation approach is trustworthy. 
We are currently hiking along that path.

\begin{suppinfo}
Basis set and frozen-core effects. Geometries used. Full list of transition energies for the benchmark Section. Additional statistical analysis. {\PopleDZ} results. Additional information for selected CI calculations.
\end{suppinfo}

\begin{acknowledgement}
D.J.~acknowledges the \emph{R\'egion des Pays de la Loire} for financial support. This research used resources of i) the GENCI-CINES/IDRIS (Grant 2016-08s015); 
ii) CCIPL (\emph{Centre de Calcul Intensif des Pays de Loire}); iii) the Troy cluster installed in Nantes; and iv) CALMIP under allocations 2018-0510 and 2018-18005 (Toulouse). 
\end{acknowledgement}

\providecommand{\latin}[1]{#1}
\makeatletter
\providecommand{\doi}
  {\begingroup\let\do\@makeother\dospecials
  \catcode`\{=1 \catcode`\}=2\doi@aux}
\providecommand{\doi@aux}[1]{\endgroup\texttt{#1}}
\makeatother
\providecommand*\mcitethebibliography{\thebibliography}
\csname @ifundefined\endcsname{endmcitethebibliography}
  {\let\endmcitethebibliography\endthebibliography}{}

\end{document}